\documentclass[aps,prb,reprint,superscriptaddress]{revtex4-1}
\usepackage{blindtext}
\usepackage{siunitx} 
\usepackage{graphicx} 
\usepackage{natbib} 
\usepackage{amsmath} 
\usepackage{amssymb}
\usepackage{ bbold }
\usepackage[utf8]{inputenc}
\usepackage{braket}
\usepackage{color}
\usepackage[table]{xcolor}
\xdefinecolor{RWTHBlue}{RGB}{0,84,159}
\xdefinecolor{myRWTHBlue}{RGB}{64,127,183}
\usepackage[
	colorlinks=true,
  	urlcolor=RWTHBlue,
  	linkcolor=RWTHBlue,
  	citecolor=RWTHBlue
  	]{hyperref}
\usepackage{units}
\usepackage[toc,page]{appendix}

\newcommand{\rfig}[1]{Fig.\,\ref{#1}}

\newcommand{\req}[1]{Eq.\,(\ref{#1})}

\newcommand{\rtab}[1]{Tab.\,\ref{#1}}

\begin{document}
\title{Transfer of a quantum state from a photonic qubit to a gate-defined quantum dot}
\author{Benjamin Joecker}
\affiliation{JARA-FIT Institute Quantum Information, Forschungszentrum J\"ulich GmbH and RWTH Aachen University, 52074 Aachen, Germany}
\affiliation{Centre for Quantum Computation and Communication Technology, School of Electrical Engineering \& Telecommunications, UNSW, Sydney, NSW, 2052, Australia}
\author{Pascal Cerfontaine}
\author{Federica Haupt}
\author{Lars R. Schreiber}
\affiliation{JARA-FIT Institute Quantum Information, Forschungszentrum J\"ulich GmbH and RWTH Aachen University, 52074 Aachen, Germany}
\author{Beata E. Kardyna\l}
\affiliation{Peter Gr\"unberg Institute, Forschungszentrum J\"ulich GmbH, 52425 J\"ulich, Germany}
\author{Hendrik Bluhm}
\email{bluhm@physik.rwth-aachen.de}
\affiliation{JARA-FIT Institute Quantum Information, Forschungszentrum J\"ulich GmbH and RWTH Aachen University, 52074 Aachen, Germany}

\begin{abstract}
Interconnecting well-functioning, scalable stationary qubits and photonic qubits could substantially advance quantum communication applications and serve to link future quantum processors. Here, we present two protocols for transferring the state of a photonic qubit  to a single-spin and to a two-spin qubit hosted in gate-defined quantum dots (GDQD). Both protocols are based on using a localized exciton as intermediary between the photonic and the spin qubit. We use effective Hamiltonian models to describe the hybrid systems formed by the the exciton and the GDQDs and apply simple but realistic noise models to analyze the viability of the proposed protocols. Using realistic parameters, we find that the protocols can be completed with a success probability ranging between 85-$\SI{97}{\percent}$. 
\end{abstract}

\maketitle

\begin{figure}[ht!]
    \includegraphics[width=0.48\textwidth]{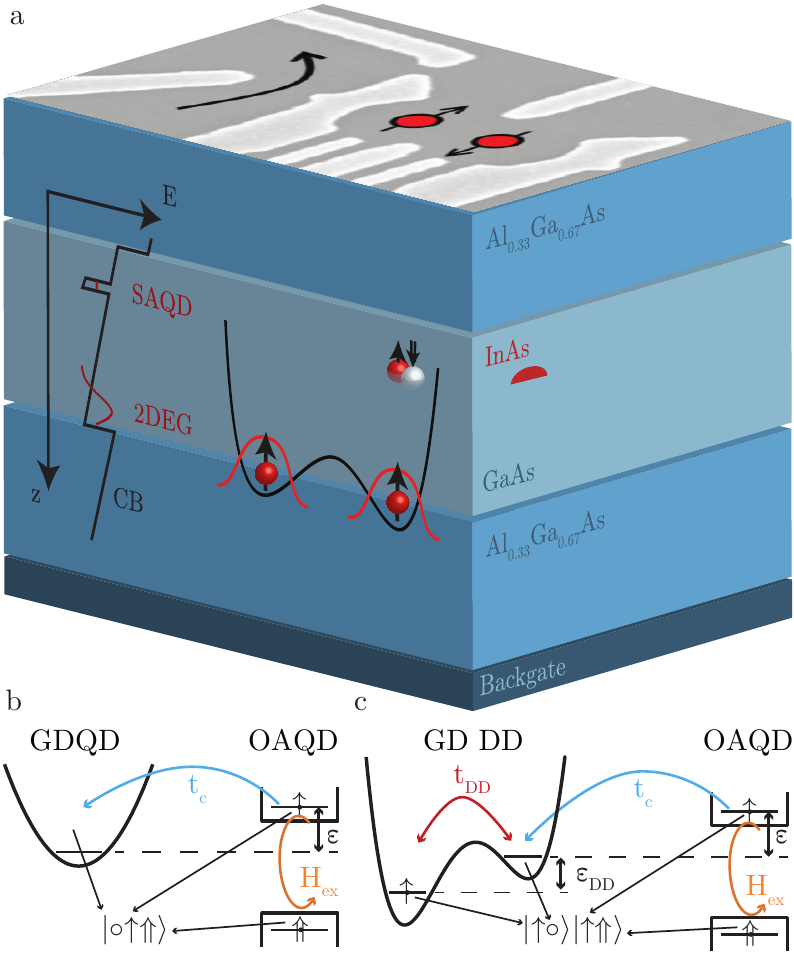}
    \caption[Schematic of the hybrid device and of the corresponding model system.]{(color online). a) Schematic of a possible heterostructure realising a hybrid device with a gate-defined double quantum dot tunnel coupled to a self-assembled quantum dot. A 2DEG emerges in the conduction band minimum at the lower $\textrm{Al}_{0.33}\textrm{Ga}_{0.67}\textrm{As}$/GaAs inverted interface. Metallic top gates can  be used to deplete the 2DEG and to create gate-defined (lateral) quantum dots. Addition of InAs during the growth of the GaAs layer leads to the formation of SAQDs.  
  b) Model of a single-level GDQD tunnel coupled to an optically active quantum dot, as discussed in Sec.~\ref{section:singleQD}. Here $t_{c}$ represents the tunnel coupling between the GDQD and the OAQD, $\varepsilon$ the energy detuning between the electronic level in the GDQD and in the exciton, and $H_{\rm ex}$ the excitonic exchange interaction. c) Model of a double dot tunnel coupled to an OAQD, as discussed in Sec.~\ref{section:DD}. In addition to the quantities defined before, $\varepsilon_{\rm DD}$ and $t_{\rm DD}$ represent the detuning and the tunnel coupling in the double dot, respectively. }
    \label{design}
\end{figure}

\section{Intoduction}
Semiconductor quantum-dot devices have demonstrated considerable potential for quantum information applications. A prominent example are gate-defined quantum dots (GDQD), i.e. quantum dots realized in semiconductor heterostructures in which individual electrons are confined by an electrostatic trapping potential. Spin qubits based on GDQD in GaAs/$\textrm{Al}_x\textrm{Ga}_{x-1}\textrm{As}$ heterostructures have 
demonstrated all key requirements for quantum information processing, such as qubit initialization, readout,\citep{Elzerman2004,Barthel2009} coherent control\citep{Koppens2006,Foletti2009} with high fidelity\citep{,Yoneda2014,Cerfontaine2016} and two-qubit gates.\citep{Nowack2011,Shulman2012} Moreover, thanks to their similarity to the transistors used in modern computer chips,  these top-down fabricated quantum dots have  good prospects for realizing large scale quantum processing nodes.  
However, unlike self-assembled quantum dots, where excellent optical control and information transfer has been demonstrated,\citep{Press2008,DeGreve2012,Gao2012} GDQDs pose  a number of challenges when it comes to couple them {\em coherently} with light.  The problems come from the lack of exciton confinement: while the electron  states are confined, the hole states are not.  Since in the creation of an exciton the spin of the photo-excited electron is always entangled with the one of the hole, discarding the hole-spin inevitably  leads to decoherence of the electron spin. This limits considerably the possibility of optically controlling and manipulating spins in GDQDs, and it hinders their applicability in quantum communications.  

Despite these difficulties, first steps towards the goal of coherently coupling photons and electron spins in GDQDs have already been made, by trapping and detecting photo-generated carriers in  GDQDs,\cite{Fujita2013} and by proving transfer of angular momentum between photons and electrons.\citep{Fujita2015} Much of this effort is motivated by the fact that robust spin-photon entanglement is a key requirement for quantum repeaters for long-distance quantum communications,\citep{Dur1999} as well as for distributed quantum computing, where different computing nodes based on GDQD are connected by optical channels.\citep{VanMeter2006}
One strategy to avoid  entanglement between the spins of the electron and the hole is to use g-factor engineering to obtain a much smaller g-factor for the electrons than for holes.\citep{Kosaka2008,Kuwahara2010,Kosaka2011} 

Here we propose a different strategy, which relies on a localized exciton in an optically active quantum dot (OAQD) as interface between a photonic qubit and a spin qubit in a GDQD. The OAQD could be a self-assembled quantum dot (SAQD) -- as also proposed by Engel and coworkers\citep{Engel2006} -- an impurity,\citep{Fu2008} or a bound exciton localized with local electric-gates by exploiting the quantum Stark effect.\citep{Schinner2013}  Using effective Hamiltonian models to describe the hybrid system formed by a bound exciton tunnel coupled to a GDQD, we analyse the feasibility of two different information transfer protocols. First,  we consider the case where the quantum state of the photon is mapped onto the state of a single-spin qubit, and then the case where the mapping is to  a singlet-triplet qubit  in a double GDQD. In both cases, the first step of the transfer process is the photo-excitation of an exciton in the OAQD in the Voigt configuration, i.e. in the presence of a strong in-plane magnetic field and normal incident light beam.  We focus in particular on the effects that can hinder the coherent transfer of the photo-excited electron from the OAQD to the GDQD, assuming 
a unitary mapping between the photon state and the exciton state. Throughout the paper we use a InAs SAQD as a concrete example of OAQD (see Fig.~\ref{design}).  The described protocols can however be straightforwardly extended to other OAQD with appropriate tunnel coupling to the GDQD. We estimate the performance of the protocols using a realistic set of parameters for InAs SAQDs.  According to these estimates, the proposed protocols could be completed with a success probability of approximately  85\% for the case of the singlet-triplet qubit, and up to 97\% probability for the single-spin qubit.

The paper is organized as follows. In Section~\ref{section:singleQD}, we discuss in detail the protocol for transferring information to a single-spin qubit, including the possible error sources (section~\ref{section:singleQD}B).  Section~\ref{section:DD} is dedicated to the  protocol for transferring information to a singlet-triplet qubit encoded in a double dot. Details on how we deal with the different noise sources  and on the model Hamiltonian employed in Sec.~\ref{section:DD} are given in Appendix~\ref{app:dephasing-noise} and~\ref{app:hamiltonians}, respectively.

\section{Information transfer to a single spin-qubit}\label{section:singleQD}
\subsection{Transfer protocol}
Transferring the information encoded in the polarization of one photon to the spin of one electron in a GDQD using an OAQD as intermediary requires
two steps: (i) the creation of a bound exciton in the OAQD by absorption of the incident photon;  (ii) the adiabatic transfer of the photo-excited electron into the GDQD.   Here, we do not model explicitly the absorption process. Rather, we assume that the process is coherent and the photo-generated exciton in the OAQD reflects the state of the absorbed photon, and discuss under which conditions the photo-excited electron can be coherently transferred to the GDQD.  

In OAQDs embedded in GaAs, the light- and the heavy-hole bands are split in energy by several tens of meV due to strain\cite{Bayer2002b} or confinement.\citep{ElKhalifi1989}  In the following we will use the notation $\ket{\downarrow\Uparrow}_{z},\ket{\uparrow\Downarrow}_{z},\dots$ to indicate the electron-hole states, where  the regular arrow 
represents the projection of the electron spin along the growth-direction $z$ ($S^{(\rm e)}_z=\pm1/2\hbar$), and double arrows 
the  projection of the heavy-hole spin ($S^{(\rm h)}_z=\pm 3/2\hbar$). In this system, electron and hole states with antiparallel spins, $\{\ket{\downarrow\Uparrow}_{z},\ket{\uparrow\Downarrow}_{z}\}$, have angular momentum $\pm\hbar$,  
and are optically addressable with circularly polarized light. Hence, they are referred to as bright states. 
States with parallel spin, $\{\ket{\uparrow\Uparrow}_{z}, \ket{\downarrow\Downarrow}_{z}\}$, are optically inactive and are referred to as dark states. The Hamiltonian of the electron-hole exchange interaction takes the block-diagonal form  
\citep{Bayer2002b,Pikus1971a,VanKesteren1990a,Ivchenko1993b,Blackwood1994a}
\begin{align}
\mathbf{H}_{0,z}=\frac{1}{2}
\begin{pmatrix}
\Delta_0 & \Delta_1 & 0 &0 \\
\Delta_1 & \Delta_0 & 0 & 0 \\
0 & 0 & -\Delta_0 & \Delta_2 \\
0 & 0  & \Delta_2 & -\Delta_0
\end{pmatrix},
\label{HSAQDZbasis}
\end{align}
with respect to the basis  $\{\ket{\downarrow\Uparrow}_{z},\ket{\uparrow\Downarrow}_{z}, \ket{\uparrow\Uparrow}_{z}, \ket{\downarrow\Downarrow}_{z}\}$. 
Here, $\Delta_0$ is the energy splitting between the dark and the bright states originating from the electron-hole exchange interaction. The off-diagonal terms $\Delta_{1}$ and $\Delta_{2}$ are responsible for the energy splitting of the bright and dark excitons, respectively. 

The excitation of the bright-states by photo-absorption induces entanglement between the spins of the electron and that of the hole as follows: $\alpha \ket{\sigma^{+}}+\beta  \ket{\sigma^{-}} \to  \alpha \ket{\downarrow\Uparrow}_{z} +\beta \ket{\uparrow\Downarrow}_{z}$, where $\alpha$ and $\beta $ are complex numbers and  $\ket{\sigma^{+}}$ and $\ket{\sigma^{+}}$ represent left and right circularly polarized photons, respectively. This poses a fundamental problem if we want to map the state of the photon onto the spin of the electron only, and discard the hole. 
To avoid this problem, it is necessary to eliminate the entanglement between the spins of the electron and of the hole. 
One  way to achieve this is via g-factor engineering.\citep{Kosaka2008,Kuwahara2010,Kosaka2011}  However, one difficulty with this approach is that the resulting small Zeeman splitting between the electron spin-states makes the system very susceptible to nuclear spin fluctuations.\citep{Reilly2010}  Furthermore, this approach is strictly limited to (Al,Ga)As-based systems and cannot be extended to other material system (e.g. II/VI)\citep{Sanaka2009,Pawlis2011}. Also, it cannot be simply extended to two-electron spin qubits with singlet-triplet encoding, which have the advantage of full electrical control.\citep{Foletti2009}

Here, we investigate a different approach, which is based on applying a strong in-plane magnetic field that mixes bright and and dark states making all of them optically accessible\citep{Bayer2002b}, as described below. We assume the in-plane magnetic field to be along the $x$-direction. Taking this as as the spin-quantization axis, the exciton Hamiltonian takes the form 
\begin{equation}\label{Hex}
\mathbf{H}_{\textrm{ex}}=\mathbf{H}_{0,x}+\mathbf{H}_{\textrm{Ze},x},
\end{equation}  
where   
\begin{widetext}
\begin{equation}
\mathbf{H}_{0,x}=\frac{1}{4}
\begin{pmatrix}
-\Delta_1-\Delta_2 & -2\Delta_0+\Delta_1-\Delta_2 & 0 & 0 \\
-2\Delta_0+\Delta_1-\Delta_2 & -\Delta_1-\Delta_2 & 0 & 0 \\
0 & 0 & \Delta_1+\Delta_2 & -2\Delta_0-\Delta_1+\Delta_2 \\
0 & 0 & -2\Delta_0-\Delta_1+\Delta_2 & \Delta_1+\Delta_2
\end{pmatrix},\label{BlockExchangeHamiltonian}
\end{equation}
represents the electron-hole exchange interaction  with respect to the basis  $\{\ket{\downarrow\Uparrow}_{x},\ket{\uparrow\Downarrow}_{x}, \ket{\uparrow\Uparrow}_{x}, \ket{\downarrow\Downarrow}_{x}\}$, and 
\begin{equation}\label{eq:HZex}
\mathbf{H}_{\textrm{Ze},x}=\frac{\mu_{\textrm{B}} B}{2}
\begin{pmatrix}
-g_{\textrm{e}}-g_{\textrm{h}} & 0 & 0 & 0\\
0 & g_{\textrm{e}}+g_{\textrm{h}} &0 & 0\\
0 & 0 & g_{\textrm{e}}-g_{\textrm{h}}& 0 \\
0 & 0 & 0 & -g_{\textrm{e}}+g_{\textrm{h}}
\end{pmatrix},
\end{equation}
\end{widetext}
accounts for the Zeeman splitting induced by the in-plane magnetic field $\vec{B}=B \hat{e}_{x}$. Here,  $g_{\rm e}$ and $g_{\rm h}$ are the g-factors for the electron and for the hole, respectively (with the sign convention that $g_{\rm e}$ is negative and $g_{\rm h}$ is positive). 
In the limit of large magnetic field ($|g_\textrm{e}\pm g_\textrm{h}| \mu_\textrm{B}B \gg \Delta_0, \Delta_1, \Delta_2$), the eigenstates of $\mathbf{H}_{\textrm{ex}}$ almost coincide with the basis kets $\{\ket{\downarrow\Uparrow}_{x},\ket{\uparrow\Downarrow}_{x}, \ket{\uparrow\Uparrow}_{x}, \ket{\downarrow\Downarrow}_{x}\}$. We will therefore denote them by their dominant basis-state contribution, e.g. $\ket{\widetilde{\uparrow\Uparrow}}_{x}= \sqrt{1-\delta}\ket{\uparrow\Uparrow}_{x}+\sqrt{\delta}\ket{\downarrow\Downarrow}_{x}$, where $\delta$ is small for large $B$.  All eigenstates have a contribution from the bright-states, i.e. they are all optically active.   
{
In general, the {\em bright-state contribution } (BC) of a state $\ket{\Psi}$ can be quantified as follow:
\begin{align}
{\rm BC}(\Psi)= \left| \braket{\Psi|\uparrow\Downarrow}_{z}\right|^2+\left| \braket{\Psi|\downarrow\Uparrow}_{z}\right|^2.
\label{BCDefinitoin}
\end{align}
The BC is a factor determining how fast a photon can be absorbed (or reemitted). 
BC substantially smaller than one are not fundamentally problematic, as they can be compensated by longer photon wave-packets.} 
Eigenstates with parallel-spins $\left(\ket{\widetilde{\uparrow\Uparrow}}_{x}, \ket{\widetilde{\downarrow\Downarrow}}_{x}\right)$  can be addressed only with horizontally polarized photons, while eigenstates with antiparallel-spins $\left( \ket{\widetilde{\downarrow\Uparrow}}_{x}, \ket{\widetilde{\uparrow\Downarrow}}_{x}\right)$ require vertically polarized photons. The idea is now to use the pair $\{\ket{\widetilde{\uparrow\Uparrow}}_{x}, \ket{\widetilde{\downarrow\Uparrow}}_{x} \}$ to map a photon state as follow: 
\begin{equation}\label{mapping-1}
\alpha \ket{\omega_{1},{\rm H}} +\beta \ket{\omega_{2}, {\rm V}} \to \alpha \ket{\widetilde{\uparrow\Uparrow}}_{x}+ \beta \ket{\widetilde{\downarrow\Uparrow}}_{x}
\end{equation}
(or, alternatively, $\alpha \ket{\omega_{1}',{\rm V}} +\beta \ket{\omega_{2}', {\rm H}} \to \alpha \ket{\widetilde{\uparrow\Downarrow}}_{x}+ \beta \ket{\widetilde{\downarrow\Downarrow}}_{x}$), where $\alpha$ and $\beta$ are complex numbers and $\ket{\omega, {\rm H (V)}}$ represents a photon state with energy $\omega$ and horizontal (vertical) polarization.  In this kind of mapping the whole information on the state of the photon is entirely mapped on the spin of the electron alone, since the excitonic states $\ket{\widetilde{\uparrow\Uparrow}}_{x}$ and $ \ket{\widetilde{\downarrow\Uparrow}}_{x} $ have the same spin projection for the hole.  

{The next step of the protocol -- and the main subject of our analysis -- is the coherent transfer of the photo-excited electron from the OAQD to a GDQD. 
If the OAQD and the GDQD are tunnel coupled, the coherent transfer between the two can be achieved by adiabatically increasing the detuning $\varepsilon$ between the electronic levels in the two system (see \rfig{design}b). Ideally, the whole transfer protocol will then work as follows: 
\begin{widetext}
\begin{equation}\label{mapping}
\alpha \ket{\omega_{1},{\rm H}} +\beta \ket{\omega_{2}, {\rm V}} \stackrel{\rm photo-excitation}{\longrightarrow} \alpha \ket{\circ\widetilde{\uparrow\Uparrow}}_{x}+ \beta \ket{\circ\widetilde{\downarrow\Uparrow}}_{x}
\stackrel{\text{ adiabatic transfer}}{\longrightarrow} 
\alpha \ket{\uparrow\circ\Uparrow}_{x}+ \beta \ket{\downarrow \circ\Uparrow}_{x}\,,
\end{equation}
\end{widetext}
where now  $\ket{\circ\widetilde{\uparrow\Uparrow}}_{x}$ represents the state where the GDQD is empty and there is an exciton with the parallel spins in the OAQD (see schematic in Fig.\ref{design}b), while $\ket{\uparrow \circ \Uparrow}_{x}$ represents the state where the electron has been transferred into the GDQD, leaving a hole alone in the OAQD (and similarly for the other states).  
}

We model the GDQD as a single electronic level and use the basis 
\begin{flalign}
\begin{split}
&\{\ket{\circ\downarrow\Uparrow}_{x}, 
\ket{\circ\uparrow\Downarrow}_{x}, 
\ket{\circ\uparrow\Uparrow}_{x}, 
\ket{\circ\downarrow\Downarrow}_{x},\\
&\ket{\downarrow\circ\Uparrow}_{x}, 
\ket{\uparrow\circ\Downarrow}_{x}, 
\ket{\uparrow\circ\Uparrow}_{x}, 
\ket{\downarrow\circ\Downarrow}_{x}\},
\label{BasisOneElectron}
\end{split}
\end{flalign} 
to represent the states of the coupled exciton-GDQD system.
With respect to this basis the Hamiltonian of the coupled exciton-GDQD system reads
\begin{align}
\mathbf{H}=\begin{pmatrix}
\mathbf{H}_{\textrm{0},x}+\mathbf{H}_{\textrm{Ze},x}+\frac{\varepsilon}{2}\mathbf{1} & t_{\textrm{c}}\mathbf{1}\\
t_{\textrm{c}}\mathbf{1} & \tilde{\mathbf{H}}_{\textrm{Ze},x} -\frac{\varepsilon}{2}\mathbf{1} 
\end{pmatrix},
\label{eq:oneElectronHamiltonian}
\end{align}
where $t_{c}$ is a spin-conserving\cite{Fujita2013} tunneling matrix element, and $\varepsilon$ is the gate-dependent energy detuning between the OAQD and the GDQD, see Fig.~\ref{design}b. We call the states where the electron and the hole are both on the OAQD {\em excitonic states}, and those where the electron has been transferred to the GDQD {\em separated states}.
The excitonic exchange-interaction, $\mathbf{H}_{{\rm 0},x}$,  has non-vanishing matrix elements only between excitonic states. 
$\tilde{\mathbf{H}}_{\textrm{Ze},x}$ has the same structure as 
in Eq.\eqref{eq:HZex}, but it can differ numerically from $\mathbf{H}_{\textrm{Ze},x}$  because of a different Zeeman splitting in the GDQD in the OAQD (e.g., because of a different $g$-factor, $\tilde{g}_{\rm e}$, in the GDQD).  
{Spin-orbit effects, which in principle can lead to spin-flip processes during the adiabatic transfer,\cite{Schreiber2011}  are not included in Eq.\eqref{eq:oneElectronHamiltonian} as we assume the dot separation to be much shorter than the spin-orbit length in GaAs,\footnote{Typical values for the spin-orbit length in GaAs are in the range $1-\SI{10}{\micro\meter}$\citep{Petta2010,Nichol2015}.}
making spin-orbit  
negligible compared to other effects. 
}
\begin{figure}
    \includegraphics[width=0.48\textwidth]{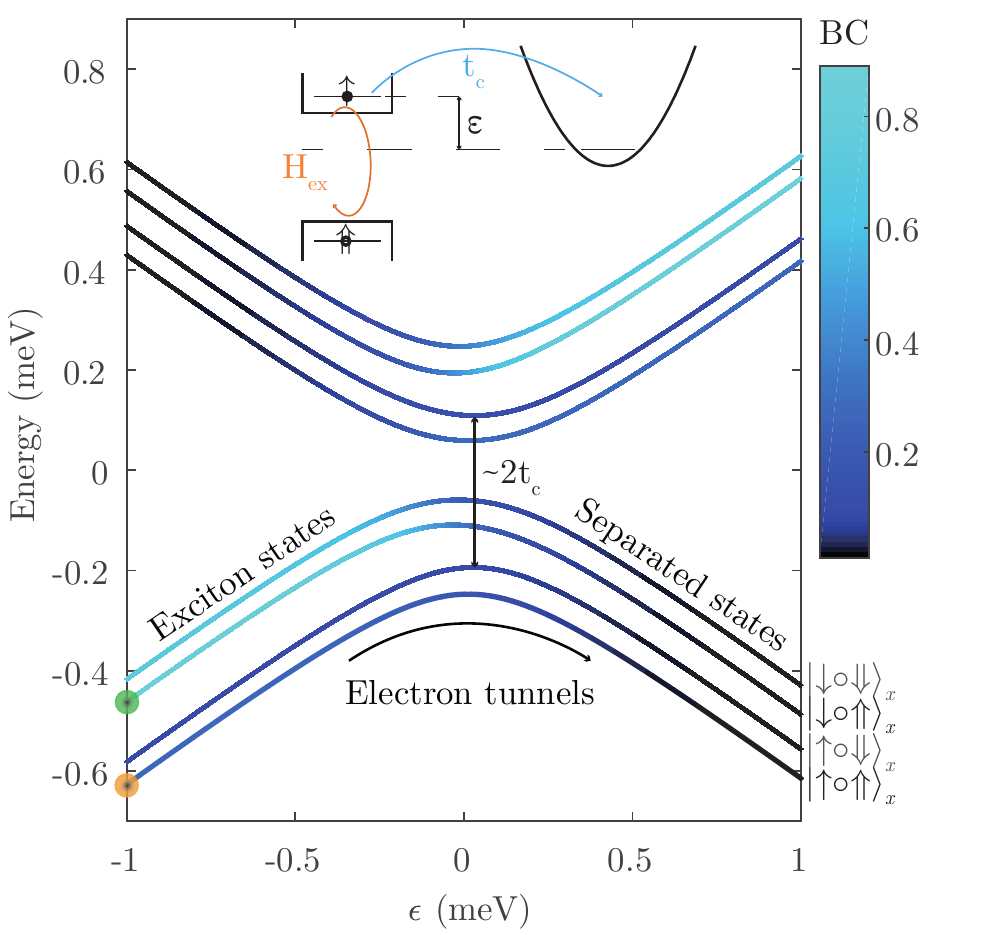}
    \caption[Schematic of the single spin protocol.]{(color online). Schematic diagram of the information-transfer process from an OAQD to a single spin qubit. Shown are the eigenenergies of the Hamiltonian \req{eq:oneElectronHamiltonian} as a function of the detuning $\varepsilon$ for the case of strong tunnel coupling ($t_\textrm{c}=\SI{150}{\micro\electronvolt}$). All remaining parameters are given in \rtab{ParameterOneElectron}.  The colour code indicates the bright-state contribution (BC) of each state, see \req{BCDefinitoin}. The bright-spots indicate the detuning at which photo-excitation occurs and the branches chosen as basis-states. Once an exciton is created in the OAQD, the photo-excited electron is transferred into the GDQD by adiabatically increasing the detuning $\varepsilon$.}
    \label{DetuningOneElecron}
\end{figure}
The eigenstates of Eq.\eqref{eq:oneElectronHamiltonian} can be easily determined numerically. In Fig.~\ref{DetuningOneElecron}  we plot the corresponding eigenenergies as a function of the detuning $\varepsilon$ for the case of large tunnel coupling $t_\textrm{c}=\SI{150}{\micro\electronvolt}$. 
This figure also includes a schematic diagram of the information-transfer process described in Eq.\eqref{mapping}.  The system is photo-excited at negative detuning, where 
excitonic states are energetically favourable (bright spots in Fig.~\ref{DetuningOneElecron}). The photo-excited electron is then transferred to the GDQD by adiabatically increasing $\varepsilon$ to the regime of separated states. The colour code  represents the  BC of each eigenstate, which clearly depends on the detuning.

\subsection{Error sources}
In practice, the viability of the protocol sketched in  Eq.\eqref{mapping}  depends on several factors and error sources. First, both selected excitonic states need  to show sufficient optical coupling. The { BC} determines how rapidly a photon can be absorbed/reemitted from a certain state, and it will therefore determine the  photon pulse-length needed for optimal absorption.  State-dependent photon pulse-shaping might be therefore required to compensate the different BC of the excitonic states. Furthermore, ideally each exciton should couple to a single photon mode, in order to reduce dissipative losses and to ensure that the photon emission/absorption process is a unitary process.  Bragg mirrors and solid immersion lenses may have to be employed to increase the collection efficiency of the OAQD.  These important, setup specific, engineering problems go however beyond the  scope of this paper.  Here we will focus instead on the intrinsic error sources that can affect the adiabatic transfer process, assuming that optimal mode engineering  ensures a unitary mapping between the photon state and the exciton state.  

The first source of errors in the adiabatic transfer process are non-adiabatic transitions to other states, which can occur if the detuning $\varepsilon$ is increased too quickly.  Given a time-dependent Hamiltonian $H(t)$, with instantaneous eigenstates $\ket{m(t)}$ (i.e. $H(t)\ket{m(t)}=E_{m}(t)\ket{m(t)}$), a necessary condition for adiabatic evolution is\citep{Messiah1962,Tong2010}
\begin{align}
\sum_{m\neq n}\left|\frac{\braket{m(t)|\dot{n}(t)}}{\omega_{mn}(t)}\right|\ll 1\label{AdiabaticCondition}
\end{align}
with $\hbar\omega_{mn}(t)=E_{m}(t)-E_{n}(t)$.  If this criterion is violated, transitions between different eigenstates are expected.
For the simple case of a two-level system, the probability of transitions between the two levels when sweeping through the avoided crossing  is given by the well known Landau-Zener formula\cite{Zener1932, Shevchenko2010}
\begin{align}
P_{\textrm{LZ}}=\exp\left(-2\pi\frac{(\Delta E_{\textrm{AC}}/2)^2}{\hbar v_\varepsilon}\right),
\label{LZF}
\end{align}  
where $\Delta E_{\textrm{AC}}$ is the energy splitting at the avoided crossing and $v_{\varepsilon}$ is the sweep speed. This formula allows to calculate the highest possible sweep speed for a given $\Delta E_{\textrm{AC}}$ and a targeted maximum transition probability (e.g., $P_{\rm LZ}=1\%$).  To obtain a similar bound on the sweep speed for the eight-level system that we are considering, we notice that the exponent in  Eq.\eqref{LZF} is closely related to the quantity on the left hand side of Eq.~\eqref{AdiabaticCondition}, being 
\begin{align}
\sum_{m\neq n}\left|\frac{\braket{m|\dot{n}}}{\omega_{mn}}\right|=
\left|\frac{\hbar\braket{1|\dot{2}}}{\Delta E_{\textrm{AC}}}\right|=\left|\frac{v_{\varepsilon}\hbar}{2\left(\Delta E_{\textrm{AC}}\right)^{2}}\right|
\end{align}
for the case of a two-level system.  This motivates us to take 
\begin{align}
\frac{1}{v_{\varepsilon}}=-\frac{4\ln(P_{\textrm{\rm LZ}})}{\pi}\sum_{m\neq n}\left|\frac{\braket{m|\frac{\partial n}{\partial\varepsilon}}}{\omega_{mn}}\right|
\label{vepsilon}
\end{align}
as a bound for the maximal sweep speed that is allowed in order to have a transition probability between different eigenstates smaller or equal to $P_{\rm LZ}$. In the following we will require $P_{\rm LZ}=1\%$, i.e. a  $99\%$ success probability for the adiabatic transfer. 

The sweep speed $v_{\varepsilon}$ and the sweep range $\Delta \varepsilon$  
determine the time $T_{\rm tr}$ on which the transfer can be completed. On the timescale of $T_{\rm tr}$ various factors that hinder the transfer process are at play. First of all, excitons can decay due to  radiative recombination. We estimate  the probability 
of recombination 
in a certain time $t$ as follows 
\begin{equation}
{P_{\textrm{rec}}(t)=1-\exp\left(-\int_{0}^{t}\Gamma(t^\prime)dt^\prime\right),}
\label{recombinationProb}
\end{equation}
where $\Gamma(t)={\rm BC}(\Psi(t))/\tau$ is the instantaneous decay rate of the state $\ket{\Psi(t)}$, with $\tau$ the characteristic lifetime of a bright exciton. 
Here $\ket{\Psi(t)}$ stands generically for the instantaneous state of the system at time $t$. Radiative recombination reduces the efficiency of the information-transfer process, but it does not introduces errors in the encoding of the information.

Other factors hindering the transfer process are charge and nuclear-spin noise, which are  well known sources of dephasing,\cite{Chirolli2008,Chekhovich2013}
causing random fluctuations of the relative phase  accumulated between two states $\ket{\Psi_1}$ and $\ket{\Psi_2}$
\begin{align}\label{phi12}
\varphi(t)=\int_{0}^{t}\frac{\Delta E_{12}(t^\prime)}{\hbar}dt^\prime 
\end{align}
 accumulated between two states $\ket{\Psi_1}$ and $\ket{\Psi_2}$, where $\Delta E_{12}$  is the energy difference between the states. Charge noise affects $\Delta E_{12}$ (and therefore $\varphi(t)$) by causing random fluctuations of the detuning $\varepsilon$. Nuclear spins in the host material affect $\Delta E_{12}$ by creating a randomly fluctuating magnetic field, the Overhauser field (OF).\citep{Barthel2009} Here we consider both quasi-static and fast uncorrelated charge noise, as well as nuclear-spin noise.
Quasi-static noise is due to fluctuations that occur on time scales much longer than the transfer time $T_{\rm tr}$, corresponding to a spectral density centered around zero frequency. We denote the root-mean-squared (rms) charge noise amplitude by $\varepsilon_{\rm rms}$. Nuclear-spin fluctuations are to a good approximation quasi-static and quantified by their rms. amplitude, $B_{\rm OF}^{(\rm rms)}$. 
On the contrary, fast uncorrelated noise has equal weight $S_{\varepsilon}$ at all frequencies (white noise).  
Typical values for $\varepsilon_{\rm rms}$, $S_{\varepsilon}$ and $B_{\rm OF}^{(\rm rms)}$ in GaAs-based devices are $\varepsilon_{\rm rms}=\SI{8}{\micro \volt} / L$,\cite{Dial2013a} $S_{\varepsilon}=\SI{5e-20}{\square\volt\per\hertz} / L^2$,\citep{Dial2013a}\footnote{The values given in Ref.\onlinecite{Dial2013a} refer to gate-equivalent voltage noises, and have to be scaled by a lever-arm of the order of $\SI{10}{\per\elementarycharge}$ to be converted into detuning energy noise.}  and $B_{\rm OF}^{(\rm rms)}=5- \SI{50}{\milli\tesla}$, where the first value is typical for GDQDs\citep{Taylor2007a} and the second for SAQDs.\citep{Taylor2007a,Prechtel2015b} L denotes the lever arm converting voltages on gates to detuning variations.
\begin{table}
\centering
\setlength{\extrarowheight}{0.1cm}
\caption{Set of realistic parameters for GaAs based devices.}
\begin{tabular}{l l l}
\hline\hline
Parameter[Source]&	Symbol&	Value\\
\hline
Dark-bright splitting\citep{Bayer2002b}& $\Delta_0$&	$\SI{100}{\micro \electronvolt}$\\
Bright state splitting\citep{Bayer2002b}&	$\Delta_1$&	$\SI{0}{\micro \electronvolt}$\\
Dark state splitting\citep{Bayer2002b}&	$\Delta_2$&	$\SI{0}{\micro \electronvolt}$\\
Magnetic field&	$B$&	$\SI{5}{\tesla}$\\
Electron g-factor&	$g_{\textrm{e}},\tilde{g}_{\textrm{e}}$&	$\SI{-0.44}{}$\\
Hole g-factor&	$g_{\textrm{h}}$&	$\SI{0.2}{}$\\
Tunnel-coupling&	$t_{\textrm{c}}$&	$50-\SI{150}{\micro \electronvolt}$\\
Exciton recombination time& $\tau$& $\SI{1}{\nano \second}$\\
Quasi static charge noise\citep{Dial2013a}& $\varepsilon_{\textrm{rms}}$& $\SI{8}{\micro\volt}/L$\\
Fast uncorrelated charge noise\citep{Dial2013a,Cerfontaine2014}& $S_\varepsilon$& $\SI{5e-20}{\square\volt\per\hertz}/L^2$\\
Gate lever-arm & $L$ &$\SI{10}{\per\elementarycharge}$\\
Nuclear spin noise SAQD\citep{Taylor2007a,Prechtel2015b} & ${B}_{\textrm{OF}}^{\textrm{(rms)}}$& $\SI{50}{\milli\tesla}$\\
Nuclear spin noise GDQD\citep{Taylor2007a}& $\tilde{B}_{\textrm{OF}}^{\textrm{(rms)}}$& $\SI{5}{\milli\tesla}$\\
Transverse e-h hyperfine ratio\citep{Prechtel2015b} & $\eta$& $\SI{<0.1}{\percent}$\\
\hline\hline
DD Parameter & & \\
\hline
DD tunnel-coupling& $t_{\textrm{DD}}$&	$\SI{50}{\micro \electronvolt}$\\
Coulomb repulsion\citep{Cerfontaine2014} & $U$& $\SI{2}{\milli\electronvolt}$\\
Coulomb energy singlet\citep{Cerfontaine2014}& $V_+$& $\SI{0.8}{\micro\electronvolt}$\\
Coulomb energy triplet\citep{Cerfontaine2014}& $V_-$& $\SI{0}{\micro\electronvolt}$
\end{tabular}
\label{ParameterOneElectron}
\end{table}
For each source of noise we evaluate the quantity $\langle \delta \varphi^{2} \rangle$ describing the dephasing due to that particular noise source, 
as detailed in Appendix~\ref{app:dephasing-noise}. Assuming that all noise sources are uncorrelated, the total dephasing is given by
$$\langle \delta \varphi^{2} \rangle_{\rm tot} =\langle \delta \varphi^{2} \rangle_{\rm ch-qs}+\langle \delta \varphi^{2} \rangle_{\rm ch-fast}+\langle \delta \varphi^{2} \rangle_{\rm spins}. $$
In order to better compare the effect of dephasing to 
other mechanisms that lead to failure of the transfer process, we introduce the 
{\em failure probability due to dephasing}, which we define as the probability of the depolarizing channel with the same average gate fidelity as the dephasing channel  (see Appendix~\ref{app:dephasing-noise})
\begin{equation}\label{eq:P_{deph-fail}}
P_{\rm deph-fail} =  \langle \delta \varphi^2 \rangle _{\rm tot}/3. 
\end{equation}

In the following, we discuss results obtained for the set of realistic parameters presented in \rtab{ParameterOneElectron}  for the case where the OAQD is an InAs self-assembled quantum dot. For simplicity we assume $\Delta_1=\Delta_2=\SI{0}{\micro\electronvolt}$, as these splittings are typically small compared to $\Delta_{0}$,\citep{Bayer2002b} and do not lead to qualitative changes. 
For the case of large tunnel coupling considered in Fig.~\ref{DetuningOneElecron} ($t_{\textrm{c}}=\SI{150}{\micro \electronvolt}$) ,  we find  that the maximal sweep velocity with $P_{\rm LZ}=\SI{1}{\percent}$ is $v_{\varepsilon}=14.8\, {\rm meV/ns}$,
so that a sweep over the whole displayed detuning range can be completed in $T_{\rm tr}=\SI{0.14}{\nano \second}$.  
{On this time scale,  the probability of radiative recombination is $P_{\rm rec}=2\%$ for the state with parallel spins and $\SI{5.5}{\percent}$ for the state with anti-parallel spins.}  At the same time, the effects of charge noise as well as of  nuclear-spin noise are rather small, giving $P_{\rm deph-fail} =\SI{0.4}{\percent}$.
{The total failure probability then is $P_{\rm fail-tot} = P_{\rm LZ} +P_{\rm rec}+P_{\rm deph-fail}\le 6.8\% $. }
Thus, the state-transfer process will work rather reliably and fail mostly because of recombination.  

\begin{figure}[ht!]
    \includegraphics[width=0.48\textwidth]{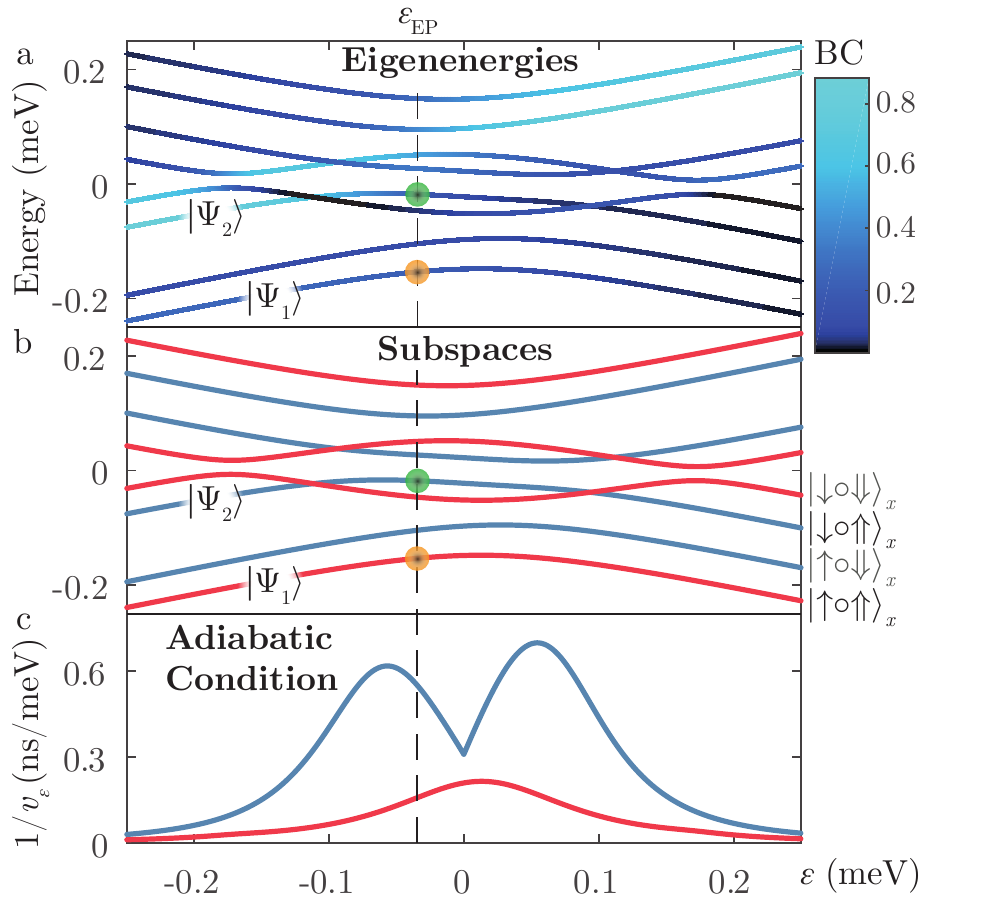}
    \includegraphics[width=0.48\textwidth]{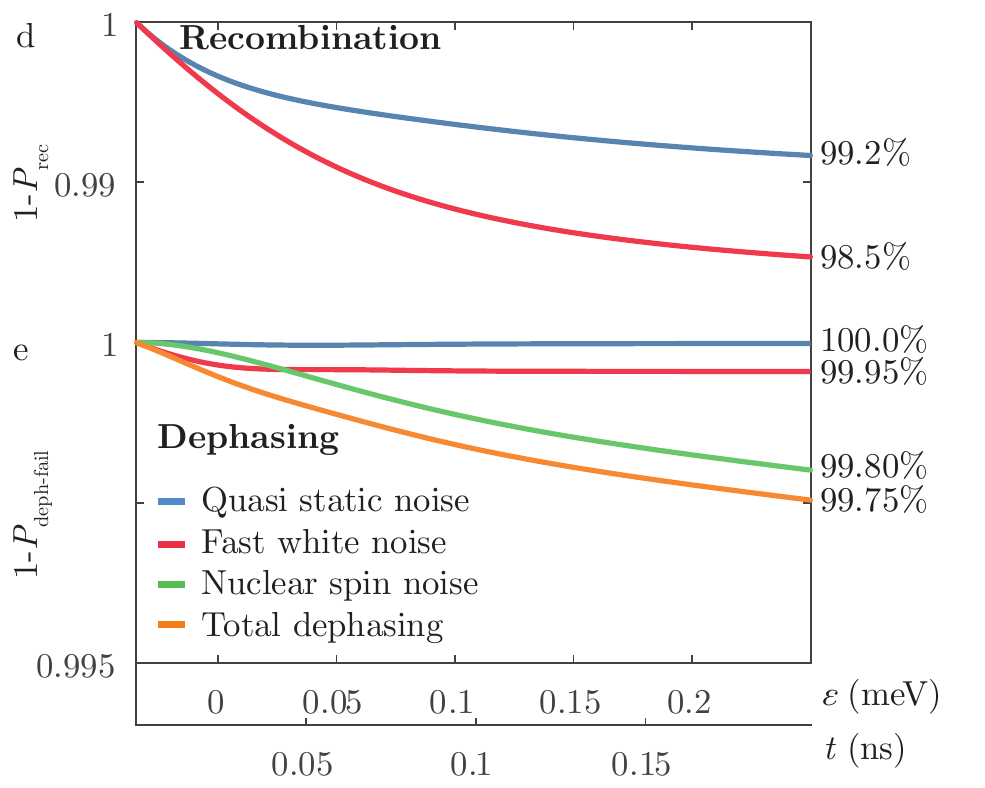}
\caption[Performance of the single spin protocol.]{(color online). Performance of the protocol for transferring information to a single spin qubits. All plots are for the parameters given in \rtab{ParameterOneElectron} with $t_\textrm{c}=\SI{50}{\micro\electronvolt}$.
 a) Eigenenergies of the coupled GDQD-exciton system, \req{eq:oneElectronHamiltonian}. The color scale indicates the BC of each state. b) Same as in a), but now the color scale represents the two independent subspaces  of the Hilbert space. The bright spots indicate the detuning $\varepsilon_{\rm EP}$ at which optical excitation occurs, as well as the branches chosen as basis-states for the information transfer-process (also labeled as $\ket{\Psi_{1}}$ and $\ket{\Psi_{2}}$).  c) Plot of the inverse maximal sweep-speed $1/v_{\varepsilon}$ that allows adiabatic evolution along the branches $\ket{\Psi_{1}}$ (red) and $\ket{\Psi_{2}}$ (blue).
d) Probability that no optical-recombination occurs during the adiabatic transfer along the branches $\ket{\Psi_{1}}$ (red) and $\ket{\Psi_{2}}$ (blue).
e) Probability of completing the adiabatic transfer without dephasing.  Different noise sources are separately accounted. The final values are displayed on the right. $t$ is the time elapsed from the beginning of the protocol.
}
\label{OneElectronResults}
\end{figure}

For smaller tunnel coupling  ($t_{\textrm{c}}=\SI{50}{\micro \electronvolt}$), the eigenstates of Eq.\eqref{eq:oneElectronHamiltonian} exhibit a series of crossing and anti-crossing, see Fig.~\ref{OneElectronResults}.  In Fig.~\ref{OneElectronResults}a the colour code represents again the bright-state contributions of the various eigenstates, while in Fig.~\ref{OneElectronResults}b red and blue indicate eigenstates with parallel or anti-parallel spins, respectively. Since these form two separate subspaces, crossings can occur between them. The labels $\ket{\Psi_1}$ and $\ket{\Psi_2}$ indicate the states involved in the adiabatic-transfer process sketched in Eq.~\eqref{mapping}, i.e. $\ket{\Psi_1}\approx\ket{\circ \uparrow \Uparrow}$, $\ket{\Psi_2}\approx\ket{\circ \downarrow \Uparrow}$  for large, negative detuning and $\ket{\Psi_1}\approx\ket{\uparrow \circ  \Uparrow}$, $\ket{\Psi_2}\approx\ket{\downarrow \circ  \Uparrow}$ for large, positive detuning. 
Fig.~\ref{OneElectronResults}c represents the inverse of the maximal sweep velocity $1/v_{\varepsilon}$ that allows adiabatic evolution along the states $\ket{\Psi_1}$ (red) and $\ket{\Psi_2}$ (blue), calculated according to Eq. \eqref{vepsilon} for $P_{\rm LZ}=1\%$.  As expected, $1/v_{\varepsilon}$ shows maxima in correspondence of the anticrossings. 
Knowing $v_{\varepsilon}$, we can calculate the total time required for the adiabatic transfer.  
For simplicity we assume that the transfer occurs with a constant speed, equal to the smaller possible maximal speed  $1/v_{\varepsilon}=\SI{0.70}{\nano\second/\milli\electronvolt}$ (see Fig.~\ref{OneElectronResults}c). Because of this low sweep speed, it is convenient to photo-excite the system not in the strongly detuned regime, but close to $\varepsilon=\SI{0}{\milli\electronvolt}$, to limit the time spent in a hybridized charge state and exposed to the strong nuclear spin field in the SAQD, as well as to minimize the probability of radiative recombination. Specifically, we 
choose photo-excitation to occur at excitation point $\varepsilon_{\rm EP}=\SI{-0.035}{\milli\electronvolt}$ (see dashed-line in \rfig{OneElectronResults}).  At this point, the two states $\ket{\Psi_1}$ and $\ket{\Psi_2}$ have the same bright-states contribution ${\rm BC}=\SI{20.3}{\percent}$. 
The transfer time required to sweep from $\varepsilon_{\rm EP}$ to  the final value $\varepsilon=0.25$\,meV at the constant speed $1/v_{\varepsilon}=\SI{0.70}{\nano\second/\milli\electronvolt}$ is $T_{\rm tr}=\SI{0.20}{\nano\second}$. 
On this timescale, radiative recombination only marginally limits the probability of a successful transfer, 
{being  $P_{\rm rec}=\SI{1.5}{\percent}$ for the state $\ket{\Psi_1}$ and $0.8\%$ for the state $\ket{\Psi_2}$ (see \rfig{OneElectronResults}d).  }
The probability of concluding the transfer without dephasing
is shown in \rfig{OneElectronResults}e.  
The results of \rfig{OneElectronResults}e correspond to a probability of failure due to dephasing of $P_{\rm deph-fail} =\SI{0.25}{\percent}$, and a total failure probability for the transfer process of $P_{\rm fail-tot} \le \SI{2.75}{\percent}$.


\section{Information transfer to a singlet-triplet qubit} \label{section:DD}
\subsection{Transfer protocol}
We consider now  a  different system, namely, instead of the coupling to a single-spin qubit, we consider the case where the electronic state in the  OAQD is tunnel coupled to a gate-defined double quantum-dot (DD), see \rfig{design}c.  One of the advantages of this setup is that a  DD can be used to encode a singlet-triplet qubit, which allows high manipulation fidelities in systems with large hyperfine interaction such as GaAs.\cite{Cerfontaine2014}
As in Sec.~\ref{section:singleQD}, we assume here that the state of a photon is mapped onto an exciton in the OAQD in the presence of an in-plane magnetic field $\vec{B}=B\vec{e}_{x}$, and  discuss under which conditions  the photo-excited electron can be coherently transfered to a neighbouring DD. Furthermore, we assume that before the optical excitation, an electron is initialized in the left side of the DD (i.e. in the one more far away from the OAQD, see Fig.~\ref{design}c) by an appropriate choice of the detuning $\varepsilon_{\textrm{DD}}$.
  
The Hamiltonian of the coupled DD-exciton system can be divided into two different subspaces: one formed by states where the both  photo-excited electron and hole are localized on the OAQD, which we call {excitonic states} (ES), and the other formed by states where the photo-excited electron has been transferred to the DD, which we call  separates states (SS). 
The subset of excitonic states is spanned by the basis $\left\{\ket{\uparrow\circ},\ket{\downarrow\circ}\right\}\otimes\left\{\ket{\downarrow\Uparrow},\ket{\uparrow\Downarrow},\ket{\uparrow\Uparrow},\ket{\downarrow\Downarrow}\right\}$), while the subset of separated states is spanned by $ \{\ket{\rm S(0,2)} \ket{\rm S(2,0)},\ket{\uparrow\downarrow},\ket{\downarrow\uparrow},\ket{\uparrow\uparrow},\ket{\downarrow\downarrow}\}\otimes\{\ket{\circ\Uparrow},\ket{\circ\Downarrow}\}$, where the spin quantization axis is taken along the direction of the applied magnetic field. In this notation, the kets on the left represent the state of the DD, with $\ket{\rm S(0,2)}$  ($\ket{\rm S(2,0)}$) representing the singlet state with the two electrons in the right  (left) side of the DD, and $\ket{\uparrow\downarrow},\ket{\downarrow\uparrow}, \dots$ representing states with one electron on each side of the DD.
The two subsets of excitonic and separated states are connected by a spin-conserving tunnel Hamiltonian, ${\bf T}$, with coupling strength $t_{c}$. The total Hamiltonian of the DD-exciton system then reads 
\begin{align}
\mathbf{H}=\begin{pmatrix}
\varepsilon\cdot\mathbf{1}+\mathbf{H}_{\textrm{ES}}&\mathbf{T}\\
\mathbf{T^\dagger}&\mathbf{H}_{\textrm{SS}}
\end{pmatrix},
\label{twoElectronHamiltonian}
\end{align}
where $\mathbf{H}_{\textrm{ES}}$ and $\mathbf{H}_{\textrm{SS}}$ are the Hamiltonians acting on the ES and SS subspaces, respectively, and $\varepsilon$ is the energy detuning between the two subspaces.   The expressions for $\mathbf{H}_{\textrm{ES}}$ and $\mathbf{H}_{\textrm{SS}}$ are given in Appendix~\ref{app:hamiltonians}. 

When occupied by two electrons, a DD can be operated as s singlet-triplet (ST) qubit, using the singlet $\ket{\textrm{S}}=\frac{1}{\sqrt{2}}\left(\ket{\uparrow\downarrow}-\ket{\downarrow\uparrow}\right)$ and triplet $\ket{\textrm{T}_0}=\frac{1}{\sqrt{2}}\left(\ket{\uparrow\downarrow}+\ket{\downarrow\uparrow}\right)$ as states  of the computational basis.\cite{Burkard1999} 
The remaining triplet states $\ket{\textrm{T}_+}
=\ket{\uparrow\uparrow}$ and $\ket{\textrm{T}_-}
=\ket{\downarrow\downarrow}$ are split off in energy  by the external magnetic field. The most relevant energy scale for the operation of a singlet-triplet  qubit is the energy splitting $J=E_{\rm S}-E_{\rm T_{0}}$ between the singlet $\ket{\rm S}$ and the triplet $\ket{\rm T_{0}}$. 
It depends on the inter-dot tunnel coupling $t_{\rm DD}$ (see \rfig{design}c), as well as the inter-dot detuning $\varepsilon_{\rm DD}$, which is an easily accessible parameter.\cite{Burkard1999,Stepanenko2011} 

A straightforward extension of the information-transfer process described in Sec.~\ref{section:singleQD} to the case of a  DD would work as follows:

\begin{align}
&\ket{\uparrow\circ}\ket{\omega_1 ,\textrm{V}} \stackrel{\text{photo-excit.}}{\longrightarrow} \ket{\uparrow\circ}\ket{\downarrow\Uparrow}
\stackrel{\text{adiabatic transf.}}{\longrightarrow}  \ket{\uparrow\downarrow}\ket{\circ\Uparrow}, 
\nonumber
\\
&\ket{\uparrow\circ}\ket{\omega_2 ,\textrm{H}}  \stackrel{{\text{photo-excit.}}}{\longrightarrow}   \ket{\uparrow\circ}\ket{\uparrow\Uparrow}
\stackrel{{\text{adiabatic transf.}}}{\longrightarrow}  \ket{\uparrow\uparrow}\ket{\circ\Uparrow}, 
\nonumber
\end{align}
representing only the evolution of the basis states).   
The state $ \ket{\uparrow\downarrow}\ket{\circ\Uparrow}$ can then be easily mapped onto the state $\ket{\rm T_{0}}\ket{\circ\Uparrow}$ by adiabatically increasing the exchange interaction in the DD (i.e. by increasing $J$).\footnote{The state $\ket{\uparrow\downarrow}\ket{\circ\Uparrow}$ has an higher energy than $ \ket{\downarrow\uparrow}\ket{\circ\Uparrow}$ because of the exchange interaction with the hole, and it therefore naturally evolves into the triplet state $\ket{\rm T_{0}}\ket{\circ\Uparrow}$  (and not into $\ket{\rm S}\ket{\circ\Uparrow}$) as the exchange interaction $J$ is adiabatically increased. The energy splitting between $\ket{\uparrow\downarrow}\ket{\circ\Uparrow}$ and $ \ket{\downarrow\uparrow}\ket{\circ\Uparrow}$ can furthermore be adjusted by introducing a small magnetic field gradient $\Delta B_{z}$ between the two dots, by dynamical  nuclear spin polarization.}
However, mapping $ \ket{\uparrow\uparrow}\ket{\circ\Uparrow}$ onto $\ket{\rm S}\ket{\circ\Uparrow}$, which is the other state of the computational basis, requires some spin-non-conserving mechanism such as, for example, the Overhauser field or the spin-orbit interaction. These effects introduce an anti-crossing between the states $\ket{\rm S}$ and $\ket{\rm T_{+}}=\ket{\uparrow\uparrow}$ in the regime where the exchange splitting $J$ approximately equals the Zeeman splitting.\cite{Stepanenko2011, Cerfontaine2014} 
This anticrossing can be used to  transform $\ket{\uparrow\uparrow}$ into $\ket{\rm S}$, however this approach would 
suffer from strong charge dephasing, since ${\textrm{T}_2^*}\propto\left(dJ/d\varepsilon_{\textrm{DD}}\right)^{-1}$ and $dJ/d\varepsilon_{\textrm{DD}}$ is fairly large at the S-$\textrm{T}_+$ transition.\citep{Dial2013a} Furthermore, the phase acquired during the process would also depend strongly on the hyperfine field. 

\begin{figure}[h]
    \includegraphics[width=0.48\textwidth]{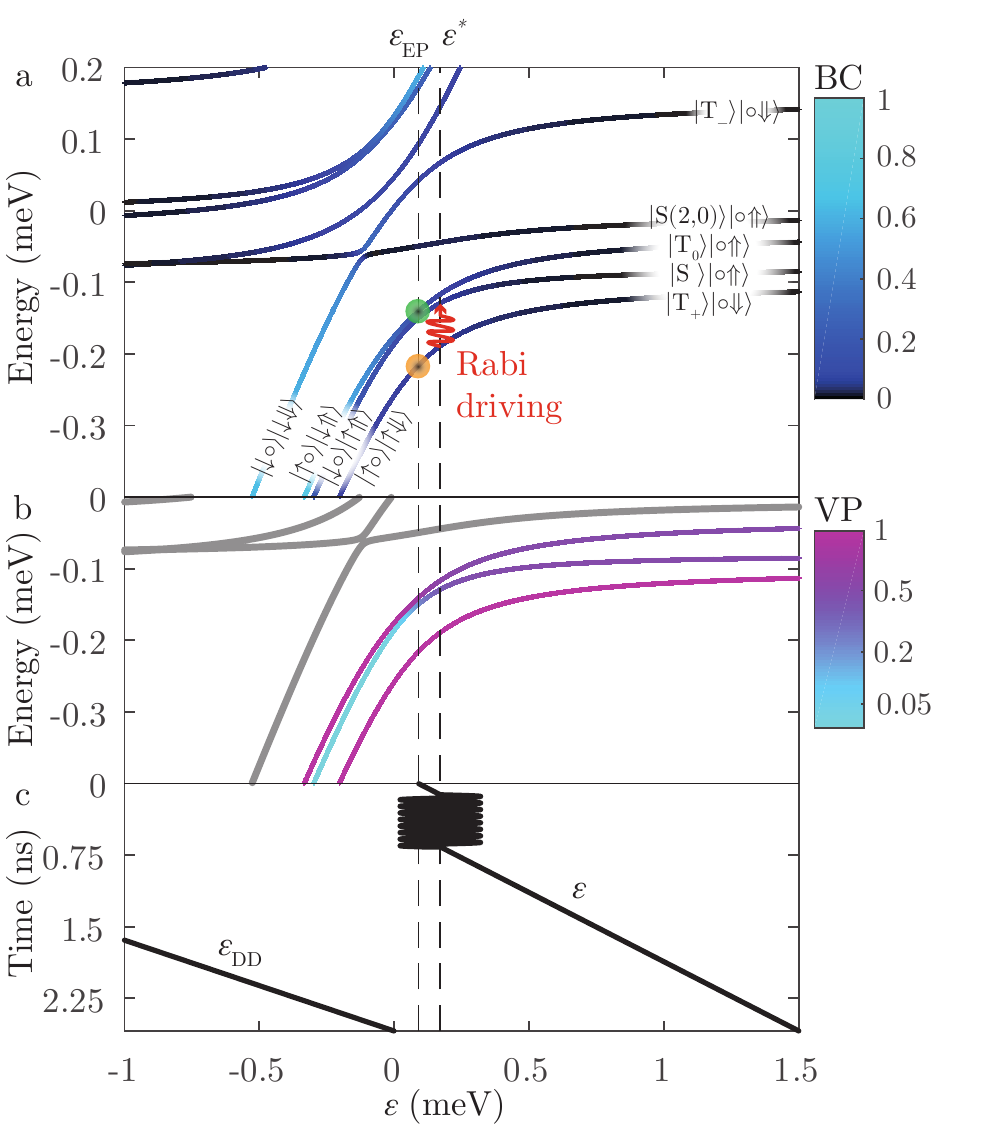}
    \caption[Schematic of the singlet-triplet transfer protocol.]{(color online). Schematic of the protocol for transferring information to a singlet-triplet qubit. a) Energy levels of the subspace sketched in \req{CouplingMechanism}, as a function of the detuning for the parameter set $\left(t_{\textrm{c}},t_{\textrm{DD}},\varepsilon_{\textrm{DD}}\right)=\left(\SI{150}{\micro\electronvolt},\SI{50}{\micro\electronvolt},-\SI{2.03}{\milli\electronvolt}\right)$. The remaining parameters are given in \rtab{ParameterOneElectron}.  The colorscale represents the bright-state contribution of the various branches. The labels close to each branch indicate the main contribution to the eigenstates in the corresponding regime. The bright spots indicate the photo-excited branches. b) Same as above, with the color scale now representing the vertical-polarization contribution of the relevant protocol branches. c) Sketch of the pulse sequence involved in the protocol. First the system is detuned to the value $\varepsilon=\varepsilon_{\rm EP}$, where the optical excitation occurs. Then $\varepsilon$ is adiabatically swept to the driving point $\varepsilon^{*}$, where a Rabi $\pi$-pulse is applied.  At the end of the pulse, $\varepsilon$ is further increased to large positive values, into the regime of separated states.  The inter-dot detuning of the double dot $\varepsilon_{\rm DD}$ is kept negative until the end of the Rabi pulse to confine the electron initialised in the DD on the left dot. At the end of the pulse, $\varepsilon_{\rm DD}$  is adiabatically swept to zero.  
    }
    \label{DetuningTwoElectron}
\end{figure}

For this reason we take a different approach, which is based on exploiting the exchange interaction between the electron initialized in the left part of the DD and the hole in the OAQD. This interaction stems from the combination of $t_{c}$, the exciton coupling in the OAQD, and the exchange interaction $J$ in the DD, according the scheme schematically represented in the following diagram:
\begin{align}
\begin{split}
\ket{\downarrow\circ}\ket{\downarrow\Downarrow}
\overset{t_{\textrm{c}}}{
\longleftrightarrow} \ket{\downarrow\downarrow}\ket{\circ\Downarrow} \\
\Delta_0''\updownarrow \quad\quad\quad\quad\quad\quad\quad\\
\ket{\downarrow\circ}\ket{\uparrow\Uparrow}
\overset{t_{\textrm{c}}}{
\longleftrightarrow} \ket{\downarrow\uparrow}\ket{\circ\Uparrow}\\
\updownarrow J\quad\quad\\
\ket{\uparrow\circ}\ket{\downarrow\Uparrow}
\overset{t_{\textrm{c}}}{
\longleftrightarrow} \ket{\uparrow\downarrow}\ket{\circ\Uparrow}\\
\Delta_0'\updownarrow \quad\quad\quad\quad\quad\quad\quad\\
\ket{\uparrow\circ}\ket{\uparrow\Downarrow}
\overset{t_{\textrm{c}}}{
\longleftrightarrow} \ket{\uparrow\uparrow}\ket{\circ\Downarrow}
\end{split}
\label{CouplingMechanism}
\end{align}
for one of the two independent subspaces that forms the Hilbert space of the system (see Appendix~\ref{app:hamiltonians}).  Here, each arrow reflects a coupling term, with  $\Delta_0'=-2\Delta_0+\Delta_1-\Delta_2$, $\Delta_0''=-2\Delta_0-\Delta_1+\Delta_2$.   The energy eigenstates corresponding to this subspace are plotted in \rfig{DetuningTwoElectron}a.
This exchange interaction between the electron initialized in the left side of the DD and the hole in OAQD creates 
an indirect coupling between the states $\ket{\downarrow\circ}\ket{\downarrow\Uparrow}$ and $\ket{\uparrow\circ}\ket{\downarrow\Downarrow}$ and
more generally, between the T$_{+}$-like and the S-like branches in \rfig{DetuningTwoElectron}a.  Exploiting this coupling, it is possible to induce transitions between these two branches by applying a suitable ac-modulation of the detuning $\varepsilon$.

This fact can be used to transfer information encoded into photons with different energy but the same polarization according to the following scheme:
\begin{align}
&\ket{\uparrow\circ}\ket{\omega_1 ,\textrm{V}} \stackrel{\rm photo-excit.}{\rightarrow} \ket{\uparrow\circ} \ket{\downarrow \Uparrow}\stackrel{\text{adiabatic transf.}}{\rightarrow} 
\ket{\textrm{T}_0}\ket{\circ\Uparrow},\nonumber \\
&\ket{\uparrow\circ}\ket{\omega_2 ,\textrm{V}} \stackrel{\rm photo-excit.}{\rightarrow} \ket{\uparrow\circ}\ket{\uparrow\Downarrow}\stackrel{\text{Rabi \!+ \!\! ad. transf.}}{\rightarrow}   \ket{\textrm{S}}\ket{\circ\Uparrow},
\label{completeProtocol}
\end{align}
where again we only represent the evolution of the basis states. The idea is the following.
First, the system is optically excited at a certain value of the detuning $\varepsilon=\varepsilon_{\rm EP}$, transferring the state of the photon into the sub-space of excitons with anti-parallel spins. 
Then $\varepsilon$ is swept to the driving point $\varepsilon^{*}$, where 
a Rabi pulse is applied to drive the transition between the T$_{+}$-like branch and the S-like branch. During the whole procedure the double-dot detuning $\varepsilon_{\rm DD}$ is set to finite negative values 
to provide a large enough $J$ and to prevent tunnelling of the electron initialized in the left part of the DD to the right part. After the Rabi pulse, $\varepsilon_{\rm DD}$ is swept to zero and $\varepsilon$ is tuned to the regime of separated states, thus mapping the state into the subspace spanned by the computational basis $\{\ket{\textrm{S}}, \ket{\textrm{T}_0}\}\otimes\ket{\circ\Uparrow}$. 
The pulse scheme for $\varepsilon$ and $\varepsilon_{\rm DD}$ required for such a protocol is sketched in \rfig{DetuningTwoElectron}c. In the discussion above we assumed that the DD is initialized in the $\ket{\uparrow\circ}$ state (which can be achieved with standard procedures), however  the protocol can be easily adapted to match the cases where the DD is initially in the state $\ket{\downarrow\circ}$ and/or to the case where the photon has horizontal polarization. For the sake of clarity, in the following we focus only on the case described  above.

\subsection{Feasibility of the transfer protocol} \label{sec:rabi-pulse}

\begin{figure}
    \includegraphics[width=0.48\textwidth]{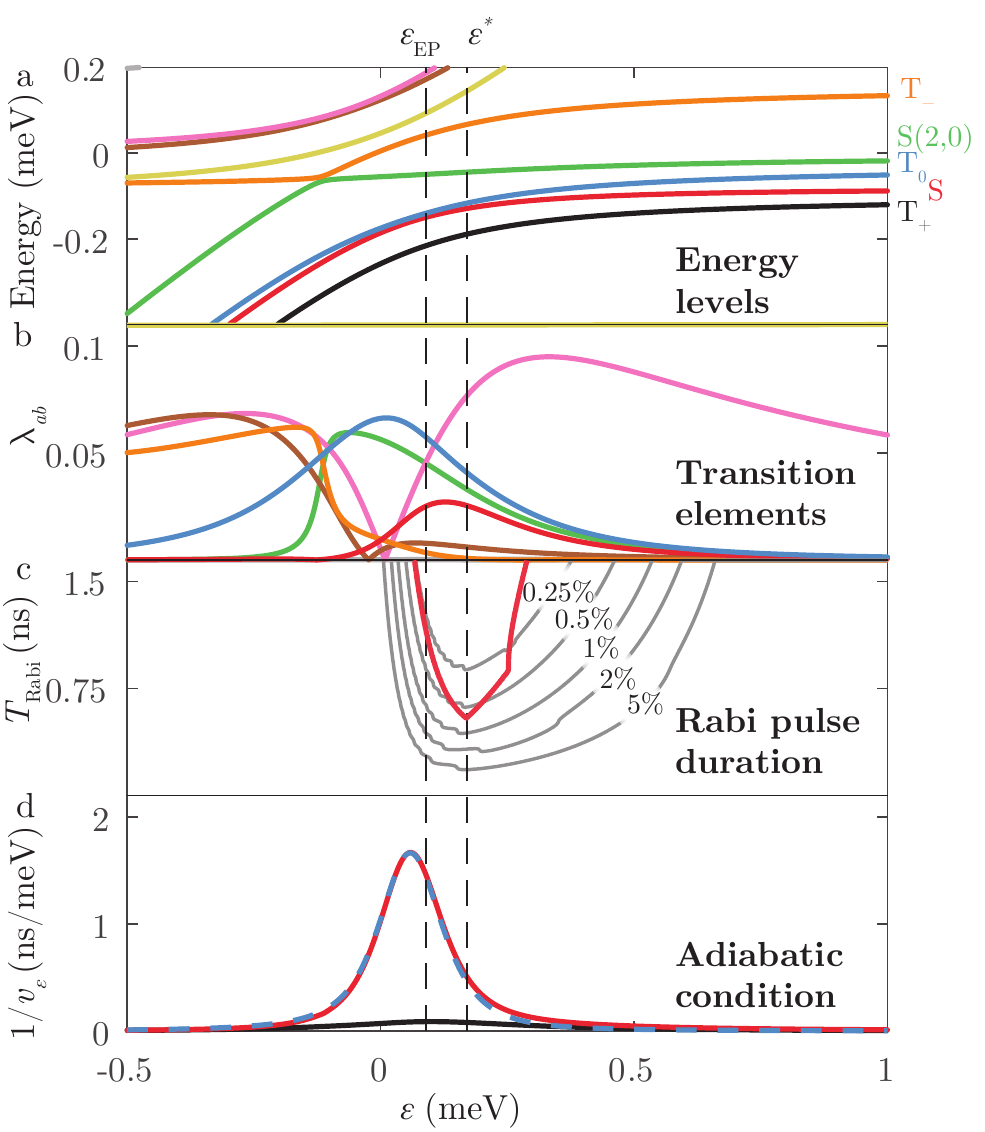}
\caption[Performance of the singlet-triplet protocol.]{(color online). Analysis of the protocol presented in \rfig{DetuningTwoElectron}. The parameter for these plots are given  in \rtab{ParameterOneElectron}, with the additional choice $\left(t_{\textrm{c}},t_{\textrm{DD}},\varepsilon_{\textrm{DD}}\right)=\left(\SI{150}{\micro\electronvolt},\SI{50}{\micro\electronvolt},-\SI{2.03}{\milli\electronvolt}\right)$. a) Energy levels of the subspace sketched in \req{CouplingMechanism}. Different branches are indicated by different colours.   
The left dashed-line marks the excitation point $\varepsilon_{\textrm{EP}}=\SI{0.09}{\milli\electronvolt}$ and the right line the driving point $\varepsilon^*=\SI{0.17}{\milli\electronvolt}$. b) Amplitude of the transition-matrix elements  $\lambda_{ab}$, from the black (T$_{+}$) branch to the  other branches. The colour code is the one defined in panel a. c) In this panel, the red curve represents the condition  $T_{\rm Rabi}={\pi \hbar}/(\Delta \varepsilon \lambda_{\rm T_{+}S})$, with $\Delta \varepsilon$ chosen such that the coupling $\lambda_{\rm T_{+}S}$ does not vary more than 50\% in the range $[\varepsilon-\Delta \varepsilon, \varepsilon+\Delta \varepsilon]$. The thin grey curves represent selected levels of constant leakage probability, $P_{\rm Rabi-leak}= P_{\Psi_{\rm T_{+}}\to \Psi_{\rm T_{0}}}=5\%$ down to $0.25\%$, calculated according to \req{eq:Pij}. 
d)  Maximal sweep-speed allowed for adiabatic transfer along the T$_{+}$-like branch (black) and along the S- and the T$_{0}$-like branches (red-blue), 
calculated according to \req{vepsilon}.
}
\label{TwoElectronResults}
\end{figure}

The transfer protocol described above depends on the choice of a number of parameters.  The excitation point $\varepsilon_{\rm EP}$ has to be chosen in such a way that the photon polarization prevents the excitation of states other than those considered in Eq.~\eqref{completeProtocol}. To do so, we consider the degree of vertical polarization of each state, i.e. the projection on the sub-space with excitonic states with antiparallel spin:
${\rm VP}(\Psi)= \left| \braket{\Psi|\uparrow\Downarrow}_{x}\right|^2+\left| \braket{\Psi|\downarrow\Uparrow}_{x}\right|^2.$ 
This quantity is plotted in Fig.~\ref{DetuningTwoElectron}b for the the relevant branches. We chose $\varepsilon_{\rm EP}$ by requiring ${\rm VP}=20\%$ for the S-like branch at the excitation point, to limit direct excitation of this branch in combination with energy selectivity, which will help to achieve a high fidelity. 

The next parameter to be fixed is the position of the driving point $\varepsilon^{*}$, which has to be chosen such as to allow an efficient Rabi $\pi$-pulse, i.e. a pulse that drives the transition between the T$_{+}$-like branch and the S-like branch in the the shortest possible time. For a two-level system driven by a rectangular pulse of the form $\varepsilon(t)=\varepsilon^{*}+\Delta\varepsilon(t)\cos(\omega_{\textrm{d}} t)$,  with $\Delta\varepsilon(t)=\Delta\varepsilon$  for time $t\in [t^{*},t^{*}+T_{\rm Rabi}]$ and zero otherwise,
the probability of transition $\ket{a}\to \ket{b}$ is given by the well-known Rabi formula \citep{Rabi1937,Boradjiev2013,Batista2015}
\begin{align} \label{eq:Prabi}
P_{a\rightarrow b}(\delta_{ab},\Omega_{ab})=\frac{\Omega_{ab}^2}{\Omega_{ab}^2+\delta_{ab}^2}\sin^2\left(\sqrt{\Omega_{ab}^2+\delta_{ab}^2} \frac{T_{\rm Rabi}}2\right),%
\end{align}
where $\delta_{ab}=\omega_{\rm d}-\omega_{ab}$, with $\hbar \omega_{ab}=E_{a}-E_{b}$, is the detuning of the driving and  $\hbar\Omega_{ab}=\Delta\varepsilon\lambda_{ab}$ the Rabi frequency, with  $\lambda_{ab}=\left.  \Braket{a|\frac{\partial H}{\partial\varepsilon}|b}\right|_{\varepsilon^{*}}$ the transition matrix element between the states.
 The conditions for a $\pi$-pulse are therefore $\delta_{ab}=0$ (resonant driving) and $T_{\rm Rabi}=\pi/{\Omega_{ab}}={\pi \hbar}/(\Delta \varepsilon \lambda_{ab})$.  For the case of the transition between the T$_{+}$-  and the S-like branches, the coupling element $\lambda_{\rm T_{+}S}$ depends on $\varepsilon$ as shown by the red-curve in  \rfig{TwoElectronResults}b. 
 For each value of $\varepsilon$, we fix the driving amplitude $\Delta \varepsilon$ by requiring that $\lambda_{\rm T_{+}S}$ does not vary more than 50\% in the detuning range $[\varepsilon-\Delta \varepsilon, \varepsilon+\Delta \varepsilon]$. The corresponding time required for a Rabi $\pi$-pulse is plotted as a red curve in \rfig{TwoElectronResults}c. The minimum of this curve gives the optimal point $\epsilon^{*}$ to apply the Rabi pulse.  For the case of \rfig{TwoElectronResults}, it is  $\varepsilon^{*}=0.17$meV, corresponding to a pulse of duration $T_{\rm Rabi}=0.55$ns and a resonance frequency $\omega_{\rm d}=2\pi\cdot 14.5$\,GHz.
 
 An important point to take into account when considering the Rabi pulse is the possible unwanted leakage to the T$_{0}$-like branch, which is energetically very close to the S-like one, and has similar coupling matrix elements.  To estimate this leakage, 
we consider  the three-level system formed by the states $\ket{\Psi_{\textrm{S}}}$, $\ket{\Psi_{\textrm{T}_+}}$ and $\ket{\Psi_{\textrm{T}_0}}$, which represent the S-like, the T$_{+}$-like  and the T$_{0}$-like branches at the excitation point $\epsilon^{*}$, and evaluate the transition probability
 \begin{align}\label{eq:Pij}
 P_{i\to j}(T_{\rm Rabi})=\left|\langle j| e^{-\frac{i}\hbar \int_{0}^{T_{\rm Rabi}} {\bf H}_{\rm RWA}(t)dt }|i\rangle \right|^{2}.
 \end{align}
Here, ${\bf H}_{\rm RWA}$ is the the Hamiltonian of the system in the rotating frame with respect to the drive with the rotating wave approximation
$$
{\bf H}_{\rm RWA}(t)=\frac\hbar2
\begin{pmatrix}
2(\omega_{\rm S T_{+}}-\omega_{\rm d}) &  \Omega_{\rm ST_{+}}(t)  &   \\
\Omega_{\rm ST_{+}}(t) & 0 &  \Omega_{\rm T_{+}T_{0}}(t)\\
 0 &  \Omega_{\rm T_{+}T_{0}}(t) & 2(\omega_{\rm T_{0}T_{+}}+\omega_{\rm d})
\end{pmatrix}.
$$
The grey curves in in Fig.~\ref{TwoElectronResults}c show selected levels of constant leakage, $P_{\rm Rabi-leak}= P_{\Psi_{\rm T_{+}}\to \Psi_{\rm T_{0}}}=5\%$ down to $0.25\%$. For the considered Rabi pulse given by the minimum of 
the red curve, the leakage probability is $P_{\rm Rabi-leak}= 0.6\%$

Another possible source of errors during the protocol are non-adiabatic transitions between the branches caused by an excessive sweep-speed. As in Sec.~\ref{section:singleQD},  we use  \req{vepsilon} to get a bound on the maximally allowed  sweep-speed, requiring $P_{\rm LZ}=1\%$. 
The (inverse) maximal sweep-speed, $1/v_{\varepsilon}$,  is plotted in  \rfig{TwoElectronResults}d. In the following we assume for simplicity a constant sweep-speed from the excitation point {$\varepsilon_{\rm EP}=0.09$~meV} to the driving point $\varepsilon^{*}=0.17$~meV, and from here to the final detuning $\varepsilon=1.5$~meV. The results of   \rfig{TwoElectronResults}d indicate that in this range of detuning the largest allowed sweep-speed is $1/v_{\varepsilon}=\SI{1.45}{\nano\second/\milli\electronvolt}$
, which then correspond to a transfer time of {0.12~ns} from the excitation point $\varepsilon_{\rm EP}$ to the driving point $\varepsilon^{*}$, and of 1.93~ns from the driving point to the final detuning $\varepsilon=1.5$~meV.

\begin{figure}
    \includegraphics[width=0.48\textwidth]{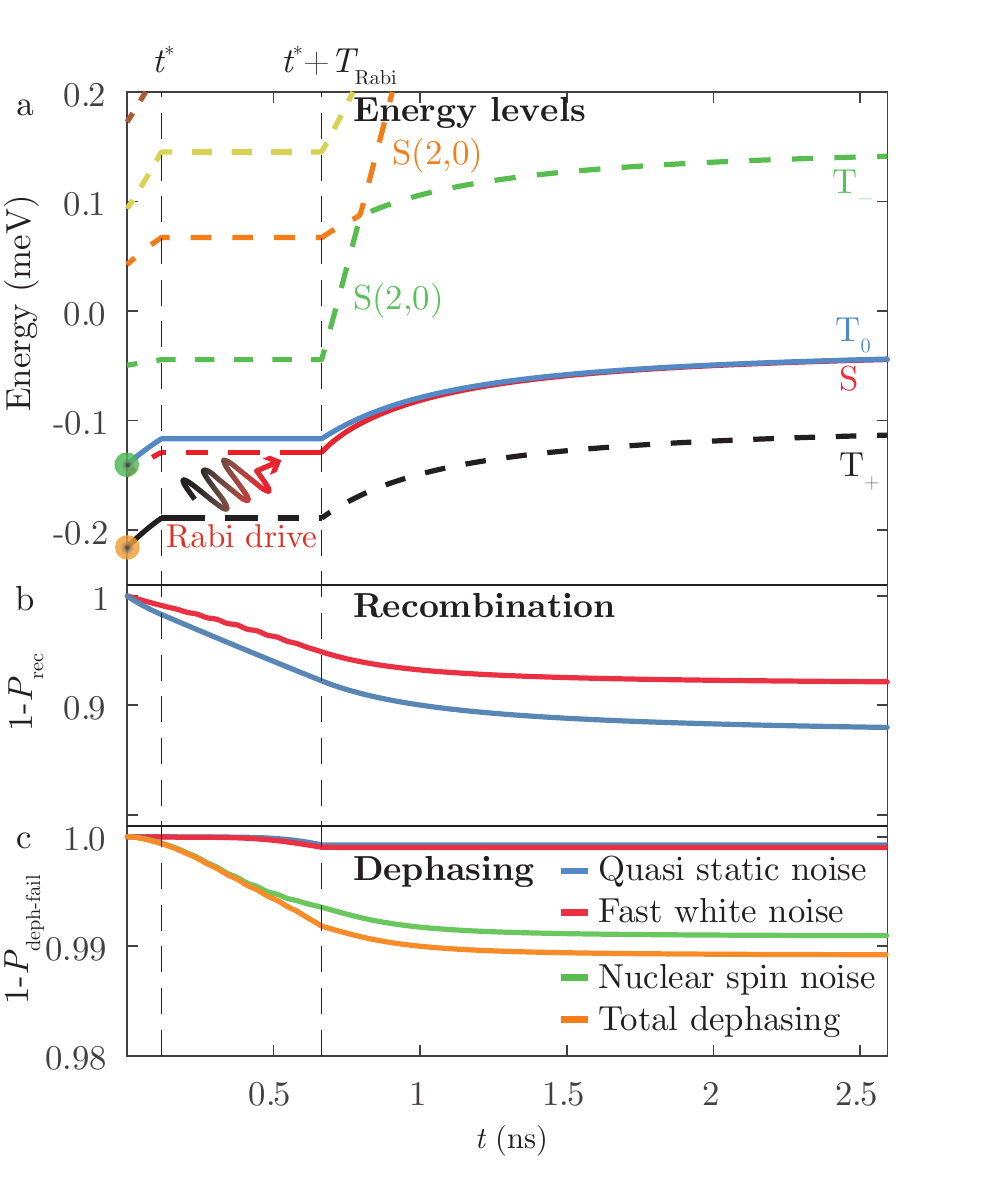}
\caption[Performance of the singlet-triplet protocol part 2.]{(color online). Performance analysis of the protocol presented in \rfig{DetuningTwoElectron}.
a) Schematic evolution of the energies of the relevant protocol branches as a function of time. Dashed lines represent unpopulated branches, while full-lines populated ones. The color code is the same as in \rfig{TwoElectronResults}a.
 b) Probability of completing the transfer without recombination for the S-like branch (red) and the $\textrm{T}_0$-like branch (blue).
  c) Failure probability due to dephasing due to different noise sources.
}
\label{TwoElectronResults2}
\end{figure}

The energies of the relevant protocol branches as a function of time are sketched in \rfig{TwoElectronResults2}a, where we also take into account that the inter-dot detuning $\varepsilon_{\rm DD}$ is swept to zero at the end of the Rabi-pulse (see \rfig{DetuningTwoElectron}), separating the S$(2,0)$-like branch from the basis states. The bright spots represents the photo-excitation of the T$_{0}$-like and the T$_{+}$-like bands.  In the ideal case of adiabatic evolution during the sweeps  and of a perfect Rabi pulse, the basis states $\ket{\Psi_{1}}$,  $\ket{\Psi_{2}}$ (where we use the notation of App.~\ref{app:dephasing-noise}) evolve as follows: $\ket{\Psi_{1}(t)}=\ket{\Psi_{\rm T_{0}}(t)}$ for any time, and
$\ket{\Psi_{2}(t)}= \ket{\Psi_{\rm T_{+}}(t)}$  for $0<t<t^{*}$, 
$$
\ket{\Psi_{2}(t)}= e^{i\omega_{\rm d}t}\cos\left(\frac{\Omega_{\rm T_{+}S}}2 t\right) \ket{\Psi_{\rm T_{+}}}_{\varepsilon^{*}} +\sin\left(\frac{\Omega_{\rm T_{+}S}}2 t\right) \ket{\Psi_{\rm S}}_{\varepsilon^{*}}
$$
for $t^{*}<t<t^{*}+T_{\rm Rabi}$ and, finally, 
$\ket{\Psi_{2}(t)}=\ket{\Psi_{\rm S}(t)}$ for $t>t^{*}+T_{\rm Rabi}$, where again $\ket{\Psi_{\textrm{T}_0}}$, $\ket{\Psi_{\textrm{T}_+}}$ and  $\ket{\Psi_{\textrm{S}}}$, represent  the T$_{0}$-like, the T$_{+}$-like  and the the S-like branches, respectively.
Assuming this time evolution, we evaluate the probability of recombination, \req{recombinationProb}, as well as the failure probability due to dephasing, \req{eq:P_{deph-fail}}, during the protocol. The results are plotted in \rfig{TwoElectronResults2}b-c. The probability of recombination lies between $P_{\rm rec}=7.9\% - 12.0\%$, while  the failure probability due to dephasing amount to {$P_{\rm deph-fail} =  1.1\% $} and it is dominated by nuclear-spin noise.

Finally, we take into account that the performance of the Rabi pulses is also affected by charge and spin noise, as they both affect the resonance condition, as well as the Rabi frequency.  We implement this by numerically calculating the average transition probability. Assuming that all noise sources are quasi-static and uncorrelated, we calculate the failure probability due to each source of noise separately. For example, fluctuations of the detuning $\varepsilon$ cause the failure probability $P_{{\rm Rabi-fail },\delta \varepsilon}=1-\braket{P_{\Psi_{\rm T_{+}}\rightarrow \Psi_{\rm S}}( \delta_{\delta\varepsilon},\Omega_{\delta\varepsilon})}_{\delta\varepsilon}$, with  $P_{\Psi_{\rm T_{+}}\rightarrow \Psi_{\rm S}}$ as given by Eq.\eqref{eq:Prabi} and  
\begin{gather*}
\delta_{\delta\varepsilon}=\frac{\partial\omega_{\rm ST_{+}}}{\partial\varepsilon}\delta\varepsilon,\\
\hbar\Omega_{\delta\varepsilon}=\Delta\varepsilon\left.\Braket{\Psi_{\textrm{S}}|\frac{\partial H}{\partial\varepsilon}|\Psi_{\textrm{T}_+}}\right|_{\varepsilon^{*}+\delta\varepsilon}.
\end{gather*}
The average  $\langle{\cdot}\rangle_{\delta \varepsilon}$ is calculated  assuming zero-mean Gaussian fluctuations of $\delta \varepsilon$.  In a similar way we calculate the failure rate of the Rabi pulse due to fluctuations in the inter-dot detuning $\delta\varepsilon_{\textrm{DD}}$, as well as in the Overhauser field in the different dots $\delta \tilde{B}^{\rm (L)}_{\textrm{OF}}$, $\delta \tilde{B}^{\rm (R)}_{\textrm{OF}}$,  and $\delta B_{\textrm{OF}}$. Finally, we sum over all these failure probabilitys to estimate the reliability of the Rabi pulse.  For the set of parameters used in Fig.~\ref{TwoElectronResults2} we obtain $P_{\rm Rabi-fail}=0.3\%$.

The failure probabilities due to the different error sources are summarized in \rtab{PerformanceSummary}. Adding all failure probabilities, we conclude that the transfer protocol can be completed with a success probability of $P_{\rm success}>85.0\%$. 

\begin{table}
\centering
\setlength{\extrarowheight}{0.1cm}
\caption{Performance of the presented protocols.}
\begin{tabular}{l l l l}
\hline\hline
 & \multicolumn{2}{l}{$\phantom{aa} $Single-spin qubit} & singlet-triplet\\
\cline{2-4}
&\rfig{DetuningOneElecron} &\rfig{OneElectronResults} & \rfig{TwoElectronResults2}
\\
\hline
$P_{\rm LZ}$& $1\%$& $1\%$& $1\%$\\
${P_{\rm rec}} $& $2-5.4~\%$& $0.8-1.5~\%$& $7.9-12.0~\%$\\
$P_{\rm deph-fail}$& $0.4\%$& $0.3\%$& $1.1\%$\\
$P_{\rm Rabi-leak}$& -& -& $0.6\%$\\
$P_{\rm Rabi-fail}$& -& -& $0.3\%$\\
\hline
$P_{\rm success}$& $>93.2\%$& $>97.2\%$& $>85.0\%$
\end{tabular}
\label{PerformanceSummary}
\end{table}

\section{Conclusions} 
We presented a feasibility analysis of two protocols for transferring information from a photonic qubit to qubits realized in GDQDs, considering both the cases of single-spin qubits, and of singlet-triplet qubits. The protocols are based on using an OAQD as interface between the photonic and the spin qubit. Our analysis is based on effective Hamiltonian models for describing the hybrid systems formed by a bound exciton in the OAQD that is tunnel coupled to a single or to a  double GDQD. We focus in particular on the error sources that can affect the transfer process. Specifically, we take into account  the influence of  the adiabatic transfer conditions, the recombination of the exciton, the decoherence due to charge and nuclear-spin noise, as well as the inaccuracy of the Rabi pulse needed for the case of information transfer to a singlet-triplet qubit.  We use simple noise models, which are expected to account for the most important effects.  

As a concrete example, we consider the case where the OAQD is realized by a InAs SAQD.  We find that for the realistic set of parameters summarized in \rtab{ParameterOneElectron}, the single-spin protocol can be completed within the coherence time with a success probability in the range $> 93.2\%$, depending on the strength of tunnel coupling between SAQD and GDQD.  The presented version of the singlet-triplet protocol can be complete with a success probability above $P_{\rm success} > 85.0\%$.
These results are based on rather conservative estimates of charge fluctuation amplitudes,\citep{Dial2013a,Cerfontaine2014} as samples with better performances have been reported.\citep{Kuhlmann2013a} Furthermore, the OF field fluctuations can be further reduced using dynamic nuclear polarization with feedback.\citep{Bluhm2010,Latta2009,Xu2009} 
We did not address specifically the important but setup specific issue of how to ensure optimal optical coupling between the photonic qubit and the OAQD. We also did not consider the additional constrains that might occur in an experimental implementation of our protocols (e.g. achievable sweep-speed, idle and pulse rise times), though the values obtained are compatible with high-end equipment. On the other hand our estimates are conservative and leave substantial room for improvement (e.g. nonlinear sweeps and pulse shaping). We thus expect that the the proposed schemes could be implemented with reasonably high success probabilities.\\


\section{Acknowledgments}
This work was supported by the Alfried Krupp von Bohlen und Halbach Foundation and the European Research Council (ERC) under the European Union's Horizon 2020 research and innovation programme (grant agreement No. 679342).  P.C. acknowledges support by Deutsche Telekom Stiftung. 


\appendix
\section{Dephasing noise}\label{app:dephasing-noise}
The relative phase $\varphi$ accumulated in a certain time $t$ between two states $\ket{\Psi_1}$ and $\ket{\Psi_2}$ is given in Eq.\eqref{phi12}, where $\Delta E_{12}$ is the energy difference between the two states. Any physical mechanism that causes random fluctuations of $\Delta E_{12}$ leads to random fluctuations $\delta \varphi $ of the relative phase,
inducing dephasing in a superposition of $\ket{\Psi_1}$ and $\ket{\Psi_2}$. 
For the case of zero-average Gaussian noise, the induced dephasing is quantified with
\begin{equation}
e^{-\langle \delta \varphi^{2}\rangle/2}
\end{equation}
where $\langle \delta \varphi^{2}\rangle$ is the variance of the phase fluctuations.

\subsection{Charge noise}
Charge noise introduces stochastic fluctuations of the detuning, $\varepsilon \mapsto \varepsilon+\delta \varepsilon $, where $\delta \varepsilon $ is a randomly fluctuating quantity, and therefore of the energy difference 
\begin{equation}
\Delta E_{12}(\varepsilon+\delta\varepsilon)\approx\Delta E_{12}(\varepsilon)+\frac{\partial\Delta E_{12}(\varepsilon)}{\partial\varepsilon}\delta\varepsilon,
\end{equation}
and in the accumulated phase 
\begin{flalign}
\label{UncorrectableRelativePhase}
\delta\varphi &=\int_{0}^{t}\frac{\delta\varepsilon(t^\prime)}{\hbar}\chi(t^\prime)dt^\prime ,
\end{flalign}
with $\chi(t)=\partial\Delta E_{12}(t)/\partial\varepsilon$. The variance of the phase fluctuations induced by charge noise can  be written as 
\begin{align}
\braket{\delta\varphi^2}=\int_{0}^{t}dt_1\int_{0}^{t}dt_2\frac{\braket{\delta\varepsilon(t_1)\delta\varepsilon(t_2)}}{\hbar^2}\chi(t_1)\chi(t_2), 
\end{align}
where the angle-bracket represents  the statistical average over all realizations of $\delta\varepsilon$. The correlation function $\braket{\delta\varepsilon(t_1)\delta\varepsilon(t_2)}$ is nothing but the Fourier transform of the noise spectral density $S_\varepsilon(\omega)$:\citep{Cywinski2008}
\begin{align}
\braket{\delta\varepsilon(t_1)\delta\varepsilon(t_2)}
=\frac{1}{2\pi}\int_{-\infty}^{\infty}d\omega\ e^{-i\omega(t_2-t_1)}\frac{ S_\varepsilon(\omega)}{2},
\end{align}
where the additional factor $\frac{1}{2}$ takes into account that we use the one-sided spectral density. 

Here we consider two types of charge noise:  quasi-static charge noise  and fast, uncorrelated charge noise. The first type represents charge fluctuations that occur on time scales much longer than the transfer time $T_{\rm tr}$, so that the charge background is essentially static during each transfer. In this case 
$$S_\varepsilon(\omega)=4\pi\varepsilon_{\textrm{rms}}^2\delta(\omega), $$
where $\varepsilon_{\textrm{rms}}$ is the root-mean-squared fluctuation in $\varepsilon$,\citep{Dial2013a} which gives for the variance of phase fluctuations 
\begin{align}
\braket{\delta\varphi^2}_{\rm ch-qs}=\frac{\varepsilon_{\textrm{rms}}^2}{\hbar^2}\left(\int_{0}^{t}
\chi(t')dt^\prime\right)^2.
\label{qscn}
\end{align}
Vice versa, fast uncorrelated noise has equal contributions at all frequencies, i.e.  $S_\varepsilon(\omega)=S_\varepsilon=\textrm{const.}$ (white noise).  In this case the variance of phase fluctuations is given by 
\begin{align}
\braket{\delta\varphi^2}_{\rm ch-fast}=\frac{S_\varepsilon}{2\hbar^2}\int_{0}^{t}\chi(t')^2dt^\prime.
\label{fucn}
\end{align}
The quantity $\chi(t)$ can be evaluated as follows. Taking into account that the detuning $\varepsilon$ enters in the Hamiltonian Eq. \eqref{eq:oneElectronHamiltonian} as: 
\begin{align}
H_\varepsilon=\frac{\varepsilon}{2}\sum_i \ket{\textrm{ES}_i}\bra{\textrm{ES}_i}-\frac{\varepsilon}{2}\sum_i\ket{\textrm{SS}_i}\bra{\textrm{SS}_i},
\end{align}
where $\ket{\textrm{ES}_i} \in \{ \ket{\circ\downarrow\Uparrow}_{x}, 
\ket{\circ\uparrow\Downarrow}_{x}, 
\ket{\circ\uparrow\Uparrow}_{x}, 
\ket{\circ\downarrow\Downarrow}_{x} \}$ and  $\ket{\textrm{SS}_i}\in\{\ket{\downarrow\circ\Uparrow}_{x}, 
\ket{\uparrow\circ\Downarrow}_{x}, 
\ket{\uparrow\circ\Uparrow}_{x}, 
\ket{\downarrow\circ\Downarrow}_{x}\}$, the quantity $\chi(t)$ becomes
\begin{flalign}
\chi(t)&=\frac{\partial\Delta E_{12}(t)}{\partial\varepsilon} \nonumber
\\&=\braket{\Psi_2(t)|\frac{\partial H_\varepsilon}{\partial\varepsilon}|\Psi_2(t)}-\braket{\Psi_1(t)|\frac{\partial H_\varepsilon}{\partial\varepsilon}|\Psi_1(t)} \nonumber\\
&=\sum_i\left|\braket{\textrm{ES}_i|\Psi_2(t)}\right|^2-\left|\braket{\textrm{ES}_i|\Psi_1(t)}\right|^2.
\end{flalign}

For the case of the singlet-triplet qubit,  one has also to take into account fluctuations in the double-dot detuning $\varepsilon_{\rm DD}$.  Assuming the fluctuations in $\varepsilon $ and $\varepsilon_{\rm DD} $ to be uncorrelated, the total variance of phase fluctuations due to charge noise is given by  
$\langle \delta \varphi^{2} \rangle_{\rm charge}=\langle \delta \varphi^{2} \rangle_{\rm ch-qs}+\langle \delta \varphi^{2} \rangle_{\rm ch-fast}+\langle \delta \varphi^{2} \rangle_{\varepsilon_{\rm DD}-\rm qs}+\langle \delta \varphi^{2} \rangle_{ \varepsilon_{\rm DD}-\rm fast}$. Here, $\langle \delta \varphi^{2} \rangle_{\varepsilon_{\rm DD}-\rm qs}$ and $ \langle \delta \varphi^{2} \rangle_{ \varepsilon_{\rm DD}-\rm fast}$ have the same structure as Eq.\eqref{qscn} and Eq.\eqref{fucn}, but with $\chi(t)$ replaced by 
\begin{align*}
\chi_{\varepsilon_{\rm DD}}(t)=\frac{\partial\Delta E_{12}(t)}{\partial\varepsilon_{\rm DD}}
&=\left|\braket{\tilde{\textrm{S}}(2,0)|\Psi_2}\right|^2-\left|\braket{\tilde{\textrm{S}}(2,0)|\Psi_1}\right|^2\\
&-\left|\braket{\tilde{\textrm{S}}(0,2)|\Psi_2}\right|^2+\left|\braket{\tilde{\textrm{S}}(0,2)|\Psi_1}\right|^2,
\end{align*}
where $\ket{\tilde{\textrm{S}}(2,0)}=\ket{\textrm{S}(2,0)}\otimes(\ket{\circ\Uparrow}+\ket{\circ\Downarrow})$, and similarly for $\ket{\tilde{\textrm{S}}(0,2)}$.

Finally we note that, in principle, one should also model the effects of $1/f$-noise. However, these can also be taken into account with reasonable accuracy by choosing the white noise level $S_{\varepsilon}$ such that it corresponds to the $1/f$-noise level in the frequency range relevant for the experiment. \\


\subsection{Nuclear-spin noise}
GaAs is a III-V semiconductor which exhibits a nuclear spin field that interacts with the spin of the charge carriers via hyperfine interaction.  Since the number of nuclear spins interacting with an electron (or a hole) is very large (typically $10^{4}$ to $10^{6}$), their effect can be conveniently described in terms of an effective  magnetic field, the Overhauser field (OF).\citep{Chirolli2008,Bluhm2011}
Here we consider only the OF component parallel to the external magnetic field (i.e. along the $x$ axis), as this is the one that most strongly affects the dynamics of the spin of the charge carriers (transverse OF components give only higher order contributions).\citep{Bluhm2011,Neder2011,Botzem2016} 
With respect to the basis Eq.\eqref{BasisOneElectron}, the effective Hamiltonian of the hyperfine interaction can then be written as 
\begin{align}
\mathbf{H}_{\rm OF}=\mu_\textrm{B} \begin{pmatrix}
B_{\textrm{OF}} (g_{\textrm{e}}{\bf S}_{x} \!+\!\eta g_{\textrm{h}}\mathbf{J}_{x})& 0\\
0 & \tilde{B}_{\rm OF}\tilde{g}_{\textrm{e}}\mathbf{S}_{x}\! +\!\eta B_{\textrm{OF}}g_{\textrm{h}}\mathbf{J}_{x} 
\end{pmatrix}.
\label{eq:OF}
\end{align}
Here, ${\bf S}_{x}$  and $\mathbf{J}_{x}$ are spin operators for  electrons and heavy holes, with eigenvalues $+\frac12$ ($-\frac12$) for states with spin-up (spin-down). Terms with a tilde  ($\tilde{B}_{\rm OF} $,  $\tilde{g}_{\textrm{e}}$) take in to account the fact that an electron can experience both different $g$-factor and  different Overhauser field in the OAQD and in the GDQD. The factor $\eta$ accounts for the different OF experienced by the electron and the hole in the OAQD.\citep{Prechtel2015b} The block-diagonal structure reflects again the separation between excitonic states and separated-states. 

As the OF fluctuates randomly in time, it causes fluctuation in the energy difference $\Delta E_{12}$. To lowest order in the fluctuations it is 
\begin{widetext}
\begin{align}
\Delta E_{12}(B_{\textrm{OF}}+\delta B_{\textrm{OF}},\tilde{B}_{\textrm{OF}}+\delta \tilde{B}_{\textrm{OF}})\approx\Delta E_{12}(B_{\textrm{OF}},\tilde{B}_{\textrm{OF}})+\frac{\partial\Delta E_{12}}{\partial B_{\textrm{OF}}}\delta B_{\textrm{OF}}+\frac{\partial\Delta E_{12}}{\partial \tilde{B}_{\textrm{OF}}}\delta \tilde{B}_{\textrm{OF}}.
\end{align}
Spin fluctuations can be considered as quasi-static noise.\citep{Barthel2009} Assuming the fluctuations  $\delta B_{\textrm{OF}}$ and $\delta \tilde{B}_{\textrm{OF}}$ to be uncorrelated, we get for the variance of the phase fluctuations induced by spin noise the following result:
\begin{equation}
\braket{\delta\varphi^2}_{\rm spins}=\frac{\mu_\textrm{B}^2\braket{\delta B_{\textrm{OF}}^2}}{\hbar^2}\bigg(\int_{0}^{t}\frac{\partial \Delta E_{12}(t)}{\partial B_{\rm OF}} \bigg)^{2}+\frac{\mu_\textrm{B}^2\braket{\delta \tilde{B}_{\textrm{OF}}^2}}{\hbar^2}\bigg(\int_{0}^{t}\frac{\partial \Delta E_{12}(t)}{\partial \tilde{B}_{\rm OF}} \bigg)^{2}.
\label{nsn}
\end{equation}
\end{widetext}
The root-mean-square fluctuations in the nuclear spin field are referred to as $B_{\textrm{OF}}^{\textrm{(rms)}}=\sqrt{\braket{\delta B_{\textrm{OF}}^2}}$ and $\tilde{B}_{\textrm{OF}}^{\textrm{(rms)}}=\sqrt{\braket{\delta \tilde{B}_{\textrm{OF}}^2}}$ in \rtab{ParameterOneElectron}. 

In the singlet-triplet qubit case we proceed analogously, but we  have to account for the OF in the two parts of the DD, i.e. we replace the term $ \tilde{B}_{\rm OF}\mathbf{S}_{x}$ in \req{eq:OF} by $ \tilde{B}_{\rm OF}^{\rm (L)}\mathbf{S}_{x}^{\rm (L)}+\tilde{B}_{\rm OF}^{\rm (R)}\mathbf{S}_{x}^{\rm (R)}$.  We furthermore assume the fluctuations in ${B}_{\rm OF}$, $ \tilde{B}_{\rm OF}^{\rm (L)}$  and $ \tilde{B}_{\rm OF}^{\rm (R)}$ to be all independent.

\subsection{Failure probability due to dephasing}
In order to compare the effects of dephasing to other effects that lead to the failure of the transfer process, we introduce the {\em failure probability due to dephasing}, 
which we define as the probability of the depolarizing channel with the same average gate fidelity.

The average gate fidelity for a two-level system can be calculated as 
\begin{equation}
\mathcal{F}(\mathcal{E}, U) = \frac{1}{2} + \frac{1}{12} \sum\limits_{k=1}^3 \mathrm{tr}\left( U \sigma_k U^{\dagger} \mathcal{E}(\sigma_k) \right), 
\end{equation}
where $\sigma_k$ are the Pauli matrices, $U$ is a quantum gate, and $\mathcal{E}$ is a trace preserving quantum operation that approximate $U$.\citep{Bowdrey2002} 
If $\mathcal{F}(\mathcal{E}, U)=1$, then $\mathcal{E}$ implements $U$ perfectly, while $\mathcal{F}(\mathcal{E}, U)<1$ indicates that $\mathcal{E}$ is a noisy (or otherwise imperfect) implementation of $U$.
With this definition, the infidelity between the phase gate $\mathcal{U}_{\delta \varphi} = \exp(-i \frac{\delta \varphi}{2} \sigma_z)$ and the identity operation becomes
\begin{equation}
1-\mathcal{F}(\mathcal{U}_{\delta \varphi} , \mathbb{1}) = \frac{1}{3} - \frac{1}{3}\cos(\delta \varphi). 
\end{equation}
If $\delta \varphi$ is a normal distributed random variable with zero mean, the expectation value of the infidelity is 
\begin{equation}
\label{eq:depol_infid}
\langle 1-\mathcal{F}(U(\delta \varphi), \mathbb{1})\rangle = \frac{1}{3} - \langle \frac{1}{3}\cos(\delta \varphi) \rangle \approx \frac{\langle \delta \varphi^2 \rangle}{6}.
\end{equation}

A depolarizing channel is defined by 
\begin{equation}
\mathcal{E}(\rho, P) = P\frac{\mathbb{1}}{2} + (1-P) \rho,
\end{equation}
where $P$ is the depolarization probability. The infidelity of the depolarizing channel is simply
\begin{equation}
1-\mathcal{F}(\mathcal{E}(\rho, P), \mathbb{1})
= P / 2.
\end{equation}
Comparing this result with \eqref{eq:depol_infid}, we define the failure probability  due to dephasing as 
\begin{equation}
P_{\rm deph-fail} = \frac{\langle \delta \varphi^2 \rangle}{3}.
\end{equation}

Here and above $\delta \varphi$ represents the phase fluctuation due to all different noise sources. Since we assume the latter to be uncorrelated, it is
$\langle \delta \varphi^{2} \rangle =\langle \delta \varphi^{2} \rangle_{\rm ch-qs}+\langle \delta \varphi^{2} \rangle_{\rm ch-fast}+\langle \delta \varphi^{2} \rangle_{\rm spins}. $

\section{Hamiltonians $\mathbf{H}_{\rm ES}$ and $\mathbf{H}_{\rm SS}$}\label{app:hamiltonians}
Here we give explicit expressions for the Hamiltonians $\mathbf{H}_{\rm ES}$ and $\mathbf{H}_{\rm SS}$ entering, Eq.\eqref{twoElectronHamiltonian}. 
The first term, $\mathbf{H}_{\rm ES}$, represents the projection of the Hamiltonian of the coupled DD-exciton system on the subspace of excitonic states 
spanned by $\left\{\ket{\uparrow\circ}_{x},\ket{\downarrow\circ}_{x}\right\}\otimes\left\{\ket{\downarrow\Uparrow}_{x},\ket{\uparrow\Downarrow}_{x},\ket{\uparrow\Uparrow}_{x},\ket{\downarrow\Downarrow}_{x}\right\}$). It is given by
\begin{equation}
\mathbf{H}_{\textrm{ES}}=\mathbf{H}_{\rm 1e}\otimes\mathbf{1}_{\rm ex}+\mathbf{1}_{\rm 1e}\otimes\mathbf{H}_{\rm ex}
\end{equation}
where 
\begin{equation}
\mathbf{H}_{\rm 1e}=\frac{\mu_{\textrm{B}} B_{x}}{2}\begin{pmatrix}
\tilde{g}_{\textrm{e}}&0\\
0&-\tilde{g}_{\textrm{e}}
\end{pmatrix}
\end{equation}
represents the Zeeman Hamiltonian of the single electron initialized in the DD, and $\mathbf{H}_{\rm ex}$ is given in Eq.\eqref{Hex}.

Similarly,  $\mathbf{H}_{\rm SS}$ represents the projection of the Hamiltonian of the coupled exciton-DD system on the subspace of separated states 
spanned by $ \{\ket{\rm S(0,2)}_{x} \ket{\rm S(2,0)}_{x},\ket{\uparrow\downarrow}_{x},\ket{\downarrow\uparrow}_{x},\ket{\uparrow\uparrow}_{x},\ket{\downarrow\downarrow}_{x}\}\otimes\{\ket{\circ\Uparrow}_{x},\ket{\circ\Downarrow}_{x}\}$, and it is given by 
\begin{equation}
\mathbf{H}_{\textrm{SS}}=\mathbf{H}_{\rm 2e}\otimes\mathbf{1}_{\rm 1h}+\mathbf{1}_{\rm 2e}\otimes\mathbf{H}_{\rm 1h}.
\end{equation}
Here 
\begin{widetext}
\begin{align}
\mathbf{H}_{\textrm{2e}}=
\begingroup
\renewcommand*{\arraystretch}{1.5}
\begin{pmatrix}
-\varepsilon_{\textrm{DD}}+U & 0 &-\frac{t_{\textrm{DD}}}{2} & \frac{t_{\textrm{DD}}}{2} & 0 & 0 \\
0 & \varepsilon_{\textrm{DD}}+U & -\frac{t_{\textrm{DD}}}{2} & \frac{t_{\textrm{DD}}}{2} & 0 & 0 \\
-\frac{t_{\textrm{DD}}}{2} & -\frac{t_{\textrm{DD}}}{2}& \frac{V_{-}+V_{+}}{2} & \frac{V_{-}-V_{+}}{2} & 0 & 0 \\
\frac{t_{\textrm{DD}}}{2} & \frac{t_{\textrm{DD}}}{2} & \frac{V_{-}-V_{+}}{2} & \frac{V_{-}+V_{+}}{2} & 0 & 0 \\
0 & 0 & 0 & 0 & V_{-}+\tilde{g}_{\textrm{e}}\mu_{\textrm{B}} B_{x} & 0 \\
0 & 0 & 0 & 0 & 0 & V_{-}-\tilde{g}_{\textrm{e}}\mu_{\textrm{B}} B_{x}
\end{pmatrix},
\endgroup
\end{align}
\end{widetext}
represents the Hamiltonian of the doubly occupied DD in the Hund-Mulliken approximation,\citep{Burkard1999,Coish2005,Stepanenko2011}  with $+(-)\varepsilon_{\textrm{DD}}$ representing the detuning energy of an electron located in the left or right side of the DD
, and $t_\textrm{DD}$ the tunnel coupling between these two sides.  $U$ is the Coulomb repulsion for two electrons located in the same dot, while  
$V_+$ and $V_-$ denote the Coulomb energy of the singlet and triplet state with one electron located in each dot, respectively. \citep{Stepanenko2011}  The Zeeman Hamiltonian of the hole remaining in the OAQD is given by $\mathbf{H}_{\rm 1h}$, which has the same structure of  $\mathbf{H}_{\rm 1e}$, but with $\tilde{g}_{\rm e}$ replaced by $-{g}_{\rm h}$. 

The Hilbert space of the coupled OAQD-DD system can be divided into two separate sub-spaces, spanned respectively by the basis states 
$\{ \ket{\downarrow\circ}\ket{\downarrow\Downarrow}$, 
$\ket{\downarrow\downarrow}\ket{\circ\Downarrow}$, 
$\ket{\downarrow\circ}\ket{\uparrow\Uparrow}$, 
$\ket{\downarrow\uparrow}\ket{\circ\Uparrow}$, $\ket{\uparrow\circ}\ket{\downarrow\Uparrow}$,
$\ket{\uparrow\downarrow}\ket{\circ\Uparrow}$, 
$\ket{\uparrow\circ}\ket{\uparrow\Downarrow}$, $\ket{\uparrow\uparrow}\ket{\circ\Downarrow}$,
$\ket{S(2,0)}\ket{\circ \Uparrow}$, $\ket{S(0,2)}\ket{\circ \Uparrow} \}$ and by
$\{ \ket{\uparrow\circ}\ket{\downarrow\Downarrow}$, 
$\ket{\uparrow\downarrow}\ket{\circ\Downarrow}$, 
$\ket{\uparrow\circ}\ket{\uparrow\Uparrow}$, 
$\ket{\uparrow\uparrow}\ket{\circ\Uparrow}$, $\ket{\downarrow\circ}\ket{\downarrow\Uparrow}$,
$\ket{\downarrow\downarrow}\ket{\circ\Uparrow}$, 
$\ket{\downarrow\circ}\ket{\uparrow\Downarrow}$, $\ket{\downarrow\uparrow}\ket{\circ\Downarrow}$,
$\ket{S(2,0)}\ket{\circ \Downarrow}$, $\ket{S(0,2)}\ket{\circ \Downarrow} \}$.

\bibliographystyle{apsrev4-1}
\bibliography{Citations} 

\end{document}